\documentclass[%
 aip,
  amsmath,amssymb,
 preprint,
]{revtex4-1}

\usepackage[utf8]{inputenc}
\usepackage[T1]{fontenc}
\usepackage{mathptmx}
\usepackage{etoolbox}
\usepackage{comment}
\usepackage[normalem]{ulem}
\usepackage{soul}
\usepackage{xcolor,cancel}

\usepackage{xcolor} 
\usepackage{graphicx}
\usepackage{dcolumn}
\usepackage{bm}

\draft 
\makeatletter
\def\@email#1#2{%
 \endgroup
 \patchcmd{\titleblock@produce}
  {\frontmatter@RRAPformat}
  {\frontmatter@RRAPformat{\produce@RRAP{*#1\href{mailto:#2}{#2}}}\frontmatter@RRAPformat}
  {}{}
}%

\usepackage{comment}
\renewenvironment{comment}{}{}

\makeatother

\usepackage[mathlines]{lineno}
  
\begin{document}

\title[Slender vortex filaments in the Boussinesq approximation]{Slender vortex filaments in the Boussinesq Approximation}

\author{Marie Rodal}
\email[]{marie.rodal@fu-berlin.de}
\affiliation{FB Mathematik und Informatik, Freie Universität Berlin, Arnimallee 6, 14195 Berlin, Germany}
\affiliation{PLECO, Department of Biology, University of Antwerp,  Universiteitsplein 1
2610 Wilrijk, Belgium}
\author{Daniel Margerit}
\affiliation{39 avenue des grands pins, 31660 Buzet-sur-Tarn, France}
\author{Rupert Klein}
\affiliation{FB Mathematik und Informatik, Freie Universität Berlin, Arnimallee 6, 14195 Berlin, Germany}

\date{\today}

\begin{abstract}
A model for the motion of slender vortex filaments is extended to include the effect of gravity. The model, initially introduced by \citet{CallegariTing1978}, is based on a matched asymptotic expansion in which the outer solution, given by the Biot-Savart law, is matched with the inner solution derived from the Navier-Stokes equations. Building on recent work by \citet{Harikrishnan2023} the Boussinesq approximation is applied such that the density variations only enter in the gravity term. However, unlike \citet{Harikrishnan2023} the density variation enters at a lower order in the asymptotic expansion, and thus has a more significant impact on the self-induced velocity of the vortex filament. In this regime, which corresponds to the regime studied by \citet{ChangSmith2018}, the effect of gravity is given by an alteration of the core constant, which couples the motion of the filament to the motion within the vortical core, in addition to a change in the compatability conditions (evolution equations) which determine the leading order azimuthal and tangential velocity fields in the vortex core. The results are used to explain certain properties of bouyant vortex rings, as well as qualitatively explore the impact of gravity on tornado type atmospheric vorticies.
\end{abstract}


\maketitle 

\section{Introduction}
Vorticies are a fundamental part of fluid dynamics, from small scale turbulence to large scale phenomena such as hurricanes. The study of vorticies and vortex motion is not only of theoretical interest, but of extreme practical importance as vorticies like hurricanes and tornados result in untold amounts of damages every year. Tornadoes, in particular, are difficult to both predict and study due to their in general sudden appearance and relatively short life span; tornadoes usually last between a few minutes to an hour depending on tornado strength. Hence, there are considerable gaps in our knowledge of the formation and general motion of vorticies of this type; for a relatively recent review on the current understanding of tornadogenesis within supercell thunderstorms we refer the reader to  \citet{Markowski2009_tornadogenesis_review} or \citet{Davies_Jones2015}. Despite the difficulties, considerable progress has been made in constructing analytical models for describing specific aspects of tornado motion; see  \citet{Lewellen1993} or \citet{Varaksin2017} for a review. \citet{Yih2007} developed a simple model based on the Bernoulli equation, and used it to demonstrate certain properties of tornado-like vorticies such as funnel broadening and an increase in the azimuthal velocity as one approaches the ground.  \citet{ShternBorissovFazle1998} used a generalized model of the planar vortex sink to explain bulge formation in tornados, where the tornado funnel suddenly expands at some height above ground.  \citet{Loper2020_1} developed a model for turbulent boundary layer flows beneath a vortex, in an attempt to gain a better understanding of the mean flow near the ground in tornado-like vorticies. The work of Loper complements earlier work by \citet{OrubaDavidsonDormy2018}, where the authors looked at eye formation in tropical cyclones.\\
We wish to approach the problem of tornado type atmospheric vorticies from a slightly different angle, employing the tools developed for the modeling of slender vortex filaments.  
While tropical cyclones can have a diameter of several hundred kilometers, tornadoes are much narrower; typically on the order of 100 meters, which, given that they can have a height of several kilometers, puts them firmly in the category of slender vortex filaments. \\
The main difficulty in the modelling of such vorticies is that the Biot-Savart law, which gives the induced velocity at any point outside of the filament, diverges when the point of evaluation is on the filament itself. To overcome this difficulty several approaches have been proposed, of which the local induction approximation (LIA) is perhaps the most well known (e.g. \citet{Batchelor_introduction_2000}). Another well known approach is the cut-off method, in which the singularity is artificially removed by introducing a lower integration limit \citep{Saffman1970}. Neither of these methods capture the dynamics of the vortex core, resulting in potentially large modelling and predictive errors \citep{KleinKnio1995}. In an attempt to overcome these shortcomings \citet{MooreSaffmann1975} developed an asymptotic method in which the velocity of the flow inside the vortical core is found using a force balance method, which in turn is combined with the velocity of the outer flow, given by the desingularized Biot-Savart law. 
We note, however, that these authors pre-assumed a spatial structure of vorticity in the vortex core which is not consistent with the Euler and Navier Stokes equations. This was noted by \citet{CallegariTing1978}, who proposed a more systematic approach based on a matched asymptotic expansion, in which the velocity field in the core is derived from the Navier-Stokes equations. In continuation of this, \citet{KleinKnio1995} succeeded in reproducing the results of \citet{CallegariTing1978}, by deriving the core velocity from the vorticity equation and Biot-Savart law directly.  
It should be noted that in all of the mentioned approaches, the fluid is assumed to be incompressible and gravitational forces are excluded. 
 For tornadoes the Mach number can generally be assumed to be small (\citet{Varaksin2017,Lewellen1993}), opening up the possibility of using different approximation methods for low Mach number flows, such as the Boussinesq, anelastic and pseudo-incompressible approximations, to simplify the problem. See {e.g. \citet{Zeytounian2003}}, \citet{OguraPhillips1962} and \citet{Durran1989} for a description of the {Boussinesq}, anelastic and pseudo-incompressible approximations, respectively. \\ Since tornadoes can have a height of several kilometers, gravity is expected to have a considerable influence on the motion within the vortical core. Unfortunately, there are a limited amount of theoretical studies on buoyant vorticies. To the best of our knowledge the only attempts at explicitly including the force of gravity in the equations of motion was presented in \citet{ChangSmith2018} and more recently in \citet{Harikrishnan2023}. \citet{ChangSmith2018} used the force balance method of Moore and Saffmann to demonstrate certain properties of bouyant vortex rings, which had already been experimentally confirmed by \citet{Turner1957}. In contrast, \citet{Harikrishnan2023} extended the asymptotic approach of \citet{CallegariTing1978} to include the force of gravity, but only for very weak gravitational forces; in the regime studied by \citet{Harikrishnan2023} gravity only appears at second order in the asymptotic expansion of the Navier Stokes equation and consequently has very little impact on the self-induced motion of the vortical core. \citet{Harikrishnan2023} showed that for closed filaments as well as for filaments which are symmetric in some streamwise direction, the effect of gravity disappears entirely when gravity only enters at second order. We wish to build on the work done by \citet{Harikrishnan2023} and further extend the analysis of \citet{CallegariTing1978} to include the force of gravity at first order. This corresponds to a regime in which gravity makes a more significant contribution to the self induced motion of the filament, and is equivalent to the regime discussed by \citet{ChangSmith2018}. We will refrain from doing the fully compressible analysis, as was done by  \citet{TingKleinKnio2007}, but rather focus on the Boussinesq regime as was also done by \citet{Harikrishnan2023}, in which the density variations only enter in the gravity term. {For an in-depth discussion of the Boussinesq approximation and associated recent advances in the study of buoyancy driven flows and hydrodynamic convection, the reader is referred to e.g. \citet{Andreev2020}, while for a discussion of the Boussinesq approximation in the context of vortex dynamics, we refer to \citet{Saffmann1993} (section 5.8) and \citet{Wu2006} (section 12.1.2)}. We also neglect compressibility effects in the background density and pressure stratifications in this first approach to the problem. We emphasize that for vorticies with significant vertical span, such as tornadoes, the Boussinesq approximation will not hold, and one should instead apply the anelastic or pseudo-incompressible approximations to accurately capture the change in pressure with increasing height. However, we still consider this the appropriate starting point, as already at this level of complexity new and interesting phenomena are expected to make an appearance. 
 \begin{figure}
    \centering
    {\frame{\includegraphics[width = 80mm]{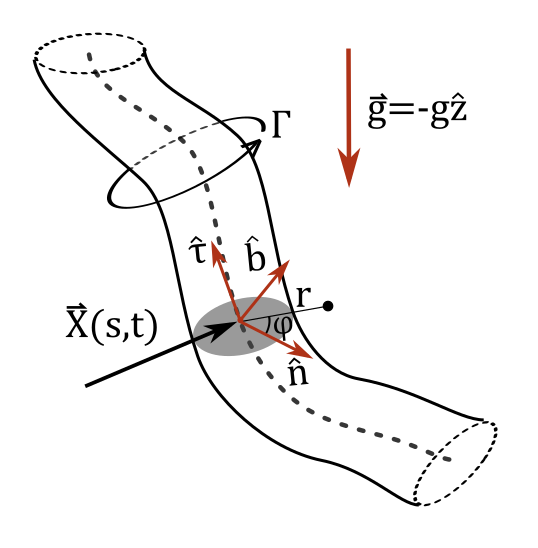}}}
    \caption{A sketch of the coordinate system.  \\}
    \label{fig:VortexSketch}
\end{figure}
Our primary objective is thus to reconstruct the results of \citet{ChangSmith2018} using a different method for estimating the flow inside the vortical core. We will restrict our analysis to finding the leading order filament velocity, and refer to  \citet{Fukumoto1991} and \citet{Margerit1996} for how to obtain the first order analysis. Our secondary objective is to shed some further light on the effect of gravity on the motion of slender vortex filaments, in particular as it relates to vorticies of a more destructive nature, noting that further analysis with different approximation methods will be necessary for an accurate description of vorticies of this type. 

\section{Outer Solution}

\label{sec:outer}
The vorticity field $\boldsymbol{\Omega}(\boldsymbol{x},t)$ of the flow to be considered here can be composed additively from the vorticity of the background flow and that of the concentrated vortex itself. By the Biot-Savart law, the 
velocity $\boldsymbol{v}(\boldsymbol{x},t)$ at any point is then composed of a velocity $\mathbf{Q}_1$ due to the self-induced motion of the vortex filament, and a velocity $\mathbf{Q}_2$ given by the background flow field. The velocity $\mathbf{Q}_1$ at any point $\boldsymbol{x}$ \textit{not too close} to the vortex core, is given by the line Biot-Savart law for an infinitely concentrated vortex:
\\
\begin{equation}
\mathbf{Q}_1(\boldsymbol{x},t) = -\frac{\Gamma}{4\pi}\int_{\mathcal{L}} \frac{\left(\boldsymbol{x}-\mathbf{X}(s',t)\right)\times \hat{\tau}(s',t) \: ds'}{\left|\boldsymbol{x}-\mathbf{X}(s',t)\right|^3}
\end{equation}
where $\mathcal{L}$ is the vortex centerline, $\mathbf{X}(s,t)$ the position vector for any point on $\mathcal{L}$, which is parametrized by the arc length parameter $s$. Here $\boldsymbol{x}$ denotes a point outside of the filament core, $\Gamma$ the total, constant circulation of the filament, and $\hat{\tau}$ the unit tangent vector.\\ For the problem at hand a curvilinear coordinate system is the natural choice. The position vector of any point $\boldsymbol{x}$ in curvilinear coordinates, can be associated to $\mathbf{X}(s,t)$ by the equation
\begin{equation}
\boldsymbol{x}(x,y,z, t) = \mathbf{X}(s,t)+r\hat{\mathbf{r}}(\varphi,s,t)
\end{equation}
Here $r$ denotes the distance from the point $\boldsymbol{x}$ to the filament $\mathbf{X}$, and $\hat{\mathbf{r}}$ is the radial unit vector for curvilinear coordinates.
See Figure 1 for an illustration of the coordinate system. Note that in this frame, we define a relative velocity $\mathbf{V}$ as $\boldsymbol{v} = \dot{\mathbf{X}}(s,t) + \mathbf{V}$.\\\\
A careful expansion of the integrand in the Biot-Savart formula leads to an expression for the behaviour of $\mathbf{Q}_1$ as the radial distance to the filament goes to zero:
\begin{eqnarray}
\label{outer_solution}
\mathbf{Q}_1(r\longrightarrow 0, \varphi, s, t) = \frac{\Gamma }{2\pi r}\hat{\boldsymbol{\theta}} + \frac{\Gamma\kappa (s,t)}{4\pi }\left[\ln\left(\frac{1}{r}\right)\right]\hat{\mathbf{b}} \hspace{1cm} \\  + \frac{\Gamma\kappa(s,t)}{4\pi}(\cos\varphi)\hat{\boldsymbol{\theta}} + \mathbf{Q}_f + \mathcal{O}\left(r \ln r\right) \nonumber 
\end{eqnarray}
where $\kappa(s,t)$ is the local curvature, $\hat{\boldsymbol{\theta}}$ the unit circumferential vector and $\hat{\mathbf{b}}$ the unit binormal vector, associated with the point $s$ on the vortex filament $\mathbf{X}(s,t)$. The vector $\mathbf{Q}_f$ is the part of the Biot-Savart integral which has a limit as $r\longrightarrow 0$. For the details on the expansion, including the exact representation of $\mathbf{Q}_f$, we refer the reader to \citet{TingKleinKnio2007} (see in particular sections 3.1.2 and 3.3.1). 
Equation (\ref{outer_solution}) cannot yield the velocity on the vortex filament itself as the expression on the right diverges as $r \longrightarrow 0$. It means that the replacement of the slender vortex by its infinitely concentrated vortex outer limit results in a singular perturbation problem: that is to say, to get a finite velocity for the slender vortex center-line, we need to account for the small but \textit{finite} thickness of the slender vortex. To get this finite velocity, we will employ a matched asymptotic expansion, closely following \citet{CallegariTing1978}, where the outer solution limit near the center-line, as given in (\ref{outer_solution}), is to be matched with the far-field limit of an inner solution derived from the Navier-Stokes equations with suitable boundary conditions on the centerline. The inner solution takes into account the finite thickness and the associated vorticity core structure which was neglected in the outer solution. We will include the force of gravity, but employ the Boussinesq approximation so that the density variations only enter in the gravitational term. {For a rigorous derivation of the Boussinesq equations from the Navier Stokes equations using asymptotic analysis the reader is referred to \citet{Zeytounian1974}.}
\section{Inner Solution}
\label{sec:inner}
In this section, we start by presenting the equations of motion for a Boussinesq fluid in filament attached (curvilinear) coordinates. This is then followed by the derivation of an equation of motion for the filament through the method of matched asymptotic expansion. The initial steps of the derivation are the same as in \citet{CallegariTing1978} and \citet{Harikrishnan2023} and are included for the sake of completeness. {Before proceeding we note the following: let us say that the core radius $\delta$ of the filament is of order $l$ and the other length scales (e.g. the radius of curvature) are of the same
order $L$. As we are considering slender filaments, the ratio $l/L$ is a small dimensionless parameter $\epsilon$.  In the following, we write dimensionless equations based on the following characteristic dimensional quantities $L$, $\Gamma$, $T_0$ and $\rho_0$, respectively for length, circulation, temperature and density. From here on, unless it is explicitly stated otherwise, all quantities and fields are dimensionless.}

\subsection{The Inner Equations} 
The Navier-Stokes equation in curvilinear coordinates is 
\begin{eqnarray}\label{IncomNavStokes}
    \ddot{\mathbf{X}} + \left(\frac{w}{h_3}-\frac{r}{h_3}\hat{\mathbf{r}}_t\cdot \hat{ \tau}\right) \dot{\mathbf{X}}_s + \frac{d\mathbf{V}}{dt} = - \frac{\mathbf{\nabla} P}{\rho} \hspace{3cm} \\ \nonumber + \frac{1}{\text{Re}}\left(\frac{1}{\rho h_3}\left(\frac{1}{h_3}\dot{\mathbf{X}}_s\right)_s + \frac{1}{\rho}\Delta \mathbf{V}\right) {- \frac{1}{\text{Fr}^2} \hat{\mathbf{z}}} \quad\quad\quad
\end{eqnarray}
where $w$  is the tangential component of the relative velocity vector $\mathbf{V}$, $\text{Re}$ and $\text{Fr}$ respectively denote the Reynolds number and Froude number{, and $-\hat{\mathbf{z}}$ is the unit gravity vector; $\mathbf{g} = - g \hat{\mathbf{z}}$ is the dimensional gravity vector with $g$ denoting the gravitational acceleration.}  The material derivative of $\mathbf{V}$ in curvilinear coordinates is given by 
\begin{equation}
\frac{d\mathbf{V}}{dt} = \frac{\partial \mathbf{V}}{\partial t} + \left(\mathbf{V} - r\frac{\partial \hat{\mathbf{r}}}{\partial t}\right) \cdot \mathbf{\nabla} \mathbf{V} 
\end{equation}
while the tangential stretching parameter, $h_3$, is
\begin{equation}
h_ 3  = \sigma[1-\kappa r\cos\varphi] = \sigma[1-\kappa r\cos(\theta + \theta_0)]
\end{equation}
where $\sigma = |\mathbf{X}_s|$.
Here $\varphi = \theta+\theta_0(s,t)$ gives the angle between the normal and radial unit vectors $\mathbf{{\hat{n}}}$ and $\mathbf{\hat{r}}$, where $\theta_0$ is included to account for the torsion $\mathcal{T}_\text{o}$ and thus makes the coordinate system orthogonal. It is easier to first derive the form of the differential operators in the $(r,\theta, s,t)$ coordinates, where the variables are orthogonal, and then obtain the form in $(r,\varphi, s,t)$ coordinates through the following transformation
\begin{equation}
\left(\frac{\partial}{\partial s} \right)_\theta = \left(\frac{\partial}{\partial s} \right)_\varphi  - \sigma \mathcal{T}_\text{o} \frac{\partial}{\partial \varphi}.
\end{equation}
In regards to the asymptotic expansion, it is simpler to use the $(r,\varphi, s,t)$ coordinates, as it avoids having to expand expressions of the form $\sin\varphi$ and $\cos{\varphi}$ where it is difficult to determine the higher order terms. A complete description of the coordinate system {including the derivation of (\ref{IncomNavStokes}), excluding gravitational forces,} is given in \citet{CallegariTing1978} (see in particular Appendix A and B), while the expansion in terms of the $(r,\varphi, s,t)$ coordinates is discussed in \citet{Margerit2002}$^,$\citep{Margerit1997_PhDthesis}. In the Boussinesq approximation the density variation only enters in the buoyancy term, $\rho \hat{\mathbf{g}}$, and can be neglected in the rest of the equation. 
This yields
\begin{eqnarray}\label{BoussinNavStokes}
\left(\ddot{\mathbf{X}} + \left(\frac{w}{h_3}-\frac{r}{h_3}\hat{\mathbf{r}}_t\cdot\hat{\tau}\right) \dot{\mathbf{X}}_s + \frac{d\mathbf{V}}{dt}\right) = - \mathbf{\nabla} P \hspace{1cm}\\ + \frac{1}{\text{Re}}\left(\frac{1}{ h_3}\left(\frac{1}{h_3}\dot{\mathbf{X}}_s\right)_s + \Delta  \mathbf{V}\right) { - \frac{\rho}{\text{Fr}^2}\hat{\mathbf{z}}}\nonumber
\end{eqnarray}
where the temperature and pressure dependent density, $\rho$, has been replaced by a constant density, except in the buoyancy term. \\
At this stage we note that for the case of a tornado-type vortex one would, at the very least, need to allow the background density to depend on the vertical coordinate $z$, in accordance with the anelastic approximation. This would, however, introduce further complications, and thus we leave this analysis for later work. \\\\
The bouyancy term can be written as 
\begin{equation}
1 + \rho'
\end{equation}
where $ \rho' = \rho - 1$ represents the density variation with respect to a reference density, and is assumed to be exclusively temperature and not pressure dependent.  \\
This can be further rewritten by noting that the variation in density will be exclusively due to temperature variations, and not pressure variations, yielding
\begin{equation}
\label{defintion_beta_T_0}
\rho - 1 = -\beta T_0(T-1)
\end{equation}
where $T_0$ denotes the reference temperature  and $\beta$ is the coefficient of thermal expansion, which in the case of an ideal gas is equal to $1/T_0$. \\\\ 
We will analyze the system in the regime  
\begin{equation}
     {\text{Re}^{-1/2} = \left(\nu/\Gamma\right)^{1/2} }= \bar{\nu}^{1/2}\epsilon
\end{equation}
where $\epsilon$ is a small, dimensionless parameter and $\bar{\nu}= \mathcal{O}(1)$ as $\epsilon\longrightarrow 0$.  
Additionally, we make the assumption that 
{
\begin{equation}
\text{Fr}^2 =  \Gamma^2/(g L^3) = \bar{\lambda}^2\epsilon^2,
\end{equation}}
where $\bar{\lambda} = \mathcal{O}(1)$, {and thus assume that the acceleration due to gravity is constant, which is a reasonable assumption for the lower atmosphere \cite{Deng2008}.} Given this distinguished limit, (\ref{BoussinNavStokes}) can be written as 
\begin{eqnarray}
\label{NS_equation_Boussniesq_approx}
    \left(\ddot{\mathbf{X}} + \left(\frac{w}{h_3}-\frac{r}{h_3}\hat{\mathbf{r}}_t\cdot\hat{\tau}\right) \dot{\mathbf{X}}_s + \frac{d\mathbf{V}}{dt}\right) = - \mathbf{\nabla} P \hspace{1.5cm}\\ + {\epsilon^2\bar{\nu}}\left(\frac{1}{ h_3}\left(\frac{1}{h_3}\dot{\mathbf{X}}_s\right)_s + \Delta  \mathbf{V}\right) -  {\epsilon^{-2}\alpha(T-1)\hat{\mathbf{z}}}\nonumber,
\end{eqnarray}
where we have defined {the dimensionless number}
\begin{equation}
\label{defintion_beta_and_z}
{\alpha = - (\beta T_0)/\bar{\lambda}^2}, 
\end{equation}
and integrated the constant buoyancy term {$-\hat{z}/(\epsilon^2\bar{\lambda}^2)$} in $-\nabla P$, as it derives from a potential. In the Boussinesq approximation, the continuity equation simplifies to it's in-compressible form, which in curvilinear coordinates $(r,\theta, s,t)$ is given by 
\begin{equation}\label{IncomDensityeq0}
\left[\left(rh_3 u\right)_{r} + \left(h_3 v\right)_\theta + r\left(w_s+\dot{\mathbf{X}}_s\cdot\hat{ \tau}\right)\right] = 0 
\end{equation}
and which in $(r,\varphi, s,t)$ coordinates becomes
\begin{equation}\label{IncomDensityeq}
\left[\left(rh_3 u\right)_{r} + \left(h_3 v\right)_\varphi + r\left(w_s - \sigma \mathcal{T}_\text{o} w_\varphi + \dot{\mathbf{X}}_s\cdot\hat{ \tau}\right)\right] = 0 
\end{equation}
where we note that, (\ref{IncomDensityeq}) corresponds to equation (2) in \citet{Margerit2002}.
\\\\
Since we have introduced a density variation to the flow, we also need an energy equation to close the system. The energy equation for a Boussinesq fluid in curvilinear coordinates, given in terms of the temperature, $T$, is 
\begin{equation}
\label{temperatureeqn}
\frac{\partial T}{\partial t} + \left(\mathbf{V} -  r\frac{\partial \hat{\mathbf{r}}}{\partial t}\right)\cdot \nabla T = \epsilon^2 \bar{\eta} \Delta T
\end{equation}
where $\bar{\eta} = \bar{\nu}/{\bar{\mu}}$ and $\bar{\mu}=\mathcal{O}(1)$ is the Prandtl number. {Note that we have made the assumption that the thermal conductivity is constant (temperature independent); the generalized case with temperature dependent thermal conductivity has been explored in e.g. \citet{LorcaBoldrini1996} \cite{LorcaBoldrini1999} and \citet{Frohlich1992}, albeit not in the context of slender vortex dynamics.} \\\\
Next, we introduce stretched radial coordinates
\begin{equation}
\bar{r} = r/\epsilon 
\end{equation}
and expand the dynamical variables in terms of $\epsilon$ as follows:

\begin{gather}
\label{expansion_u}
u(\bar{r}, \varphi, s,t ; \epsilon)  = u^{(1)}(\bar{r}, \varphi, s,t ) + \epsilon u^{(2)}(\bar{r}, \varphi, s,t ) + \cdots \\
\label{expansion_v}
v(\bar{r}, \varphi, s,t ; \epsilon) = \epsilon^{-1}v^{(0)}(\bar{r}, \varphi, s,t ) +  v^{(1)}(\bar{r}, \varphi, s,t ) + \cdots \\
\label{expansion_w}
w(\bar{r}, \varphi, s,t ; \epsilon) =\epsilon^{-1}w^{(0)}(\bar{r}, \varphi, s,t ) +  w^{(1)}(\bar{r}, \varphi, s,t ) + \cdots \qquad\\
\mathbf{X}(s,t;\epsilon) = \mathbf{X}^{(0)}(s,t) + \epsilon\mathbf{X}^{(1)}(s,t)  + \cdots \qquad
\end{gather}
Here $u, v$ and $w$ denote the radial, circumferential and tangential components of the velocity vector $\mathbf{V}$, respectively. Furthermore, to obtain non-trivial velocities $v^{(0)}$ and $w^{(0)}$, we must have that 
\begin{equation}
    P(\bar{r}, \varphi, s,t ; \epsilon)  =  \epsilon^{-2}P^{(0)}(\bar{r}, \varphi, s,t) + \epsilon^{-1} P^{(1)}(\bar{r}, \varphi, s,t ) + \cdots \\
\end{equation}
The perturbation temperature $\tilde{T} = T-1$  is expanded in an asymptotic series as follows 
\begin{equation}
\tilde{T}(\bar{r},\varphi,s,t;\epsilon) = \tilde{T}^{(0)}(\bar{r},\varphi,s,t) + \epsilon \tilde{T}^{(1)}(\bar{r},\varphi,s,t) + \cdots
\end{equation}
In addition, we require that
\begin{equation}
    \dot{\mathbf{X}}\cdot\hat{ \tau} = 0
\end{equation}
so that the filament centerline $\mathcal{L}$ forms a material curve. \\
The geometric parameters $\sigma$, $\kappa$ and $h_3$ are, through the Serret-Frenet formulas {(see (A.1) in Appendix A  of \citet{CallegariTing1978})}, functions of $\mathbf{X}(s,t)$, and thus expanded as follows:
\begin{eqnarray}
\sigma(s,t;\epsilon) &=& \sigma^{(0)}(s,t) + \epsilon\sigma^{(1)}(s,t) + \cdots \\ &=& |\mathbf{X}^{(0)}_s| +\epsilon\frac{\mathbf{X}^{(0)}_s\cdot\mathbf{X}^{(1)}_s}{|\mathbf{X}^{(0)}_s|} + \cdots \nonumber \\
\kappa(s,t;\epsilon) &=& \kappa^{(0)}(s,t) + \epsilon \kappa^{(1)}(s,t) +\cdots \\
\mathcal{T}_\text{o}(s,t;\epsilon) &=& \mathcal{T}_\text{o}^{(0)}(s,t) + \epsilon \mathcal{T}_\text{o}^{(1)}(s,t) +\cdots \\
h_3(\bar{r}, \varphi, s,t ; \epsilon) &=& \sigma^{(0)}(s,t) + \epsilon h_3^{(1)}(\bar{r}, \varphi, s,t ; \epsilon) + \cdots \\ \nonumber &=& \sigma^{(0)} + \epsilon\left[\sigma^{(1)} - \sigma^{(0)}\kappa^{(0)}\bar{r}\cos\varphi\right] + \cdots
\end{eqnarray}
Let $z_{r}$, $z_{\theta}$ and $z_\text{t}$  denote the radial, circumferential and tangential components of the constant vector $\hat{\mathbf{z}}$, such that  $\hat{\mathbf{z}}=  z_{r} \hat{\mathbf{r}} + z_{\theta} \hat{\boldsymbol{\theta}} + z_\text{t} \hat{\boldsymbol{\tau}}$. Equivalently, we can define $\hat{\mathbf{z}}= z_\text{t} \hat{\boldsymbol{\tau}} + z_\text{n} \hat{\mathbf{n}} + z_\text{b} \hat{\mathbf{b}}$. The expansion of the components of $\mathbf{\hat{z}}$ is thus given as:
\begin{gather}
\label{expansion_z_r}
z_{r}(\varphi, s,t ; \epsilon)  = z_{r}^{(0)}(\varphi, s,t ) + \epsilon z_{r}^{(1)}(\varphi, s,t ) + \cdots \nonumber \\
\label{expansion_z_theta}
z_{\theta}(\varphi, s,t ; \epsilon)  = z_{\theta}^{(0)}(\varphi, s,t ) + \epsilon z_{\theta}^{(1)}(\varphi, s,t ) + \cdots \nonumber\\
\label{expansion_z_tau}
z_\text{t}(s,t ; \epsilon)  = z_\text{t}^{(0)}(s,t ) + \epsilon z_\text{t}^{(1)}( s,t ) + \cdots\nonumber \\ \nonumber
\label{expansion_z_n}
z_\text{n}(s,t ; \epsilon)  = z_\text{n}^{(0)}(s,t ) + \epsilon z_\text{n}^{(1)}( s,t ) + \cdots\nonumber \\ \nonumber
\label{expansion_z_b}
z_\text{b}(s,t ; \epsilon)  = z_\text{b}^{(0)}(s,t ) + \epsilon z_\text{b}^{(1)}( s,t ) + \cdots\nonumber \\ \nonumber
\end{gather}
Given our choice of $\epsilon$, the viscosity terms will only enter at second order. One can show that the term
\begin{equation}
\left(\frac{1}{h_3}\left(\frac{1}{h_3}\dot{\mathbf{X}}_s\right)_s\right)
\end{equation}
is $\mathcal{O}(1)$, so this term only enters much later in the expansion.  
\\\\ 
Given this, the leading order momentum equations in the Boussinesq regime are 
\begin{subequations}
\begin{eqnarray}
\label{momentleadingorder_s}
\frac{w_\varphi^{(0)}v^{(0)}}{\bar{r}}&=& 0 \\
\label{momentleadingorder_theta}
\frac{v^{(0)}v_\varphi^{(0)}}{\bar{r}}&=& -\frac{1}{\bar{r}}P_\varphi^{(0)} \\
\label{momentleadingorder_r}
\frac{\left(v^{(0)}\right)^2}{\bar{r}}&=& P_{\bar{r}}^{(0)} 
\end{eqnarray}
\end{subequations}
while the leading order continuity and temperature equations are 
\begin{equation}
\label{contleadingorder}
    v_\varphi^{(0)} = 0
\end{equation}
and
\begin{equation}
\label{leadingordertemperature}
    T_\varphi^{(0)} = 0
\end{equation}
Combining (\ref{momentleadingorder_s}), (\ref{momentleadingorder_theta}), (\ref{contleadingorder}) and (\ref{leadingordertemperature}) we conclude that all the leading order field components are independent of $\varphi$. \\
Thus, dropping all the terms containing leading order $\varphi$-derivatives, the first order momentum equations are \\ 
\begin{subequations}
\begin{gather}
\label{momentfirstorder_s}  
\Bigg( u^{(1)}w_{\bar{r}}^{(0)}  + \frac{w_\varphi^{(1)}}{\bar{r}}v^{(0)} + \frac{w^{(0)}w^{(0)}_s}{\sigma^{(0)}} \hspace{2.5cm} \\\nonumber + w^{(0)}\kappa^{(0)}v^{(0)}\sin\varphi\Bigg)    = - \frac{P_s^{(0)}}{\sigma^{(0)}} - \alpha \tilde{T}^{(0)}z_\text{t}^{(0)} \\
\label{momentfirstorder_theta}  
\Bigg(u^{(1)} v_{\bar{r}}^{(0)} + \frac{v^{(0)}v_\varphi^{(1)}}{\bar{r}} + \frac{v^{(0)}u^{(1)}}{\bar{r}} +\frac{w^{(0)}v^{(0)}_s}{\sigma^{(0)}} \hspace{1.5cm} \\\nonumber -\left(w^{(0)}\right)^2\kappa^{(0)}\sin\varphi\Bigg)  = - \frac{P_\varphi^{(1)}}{\bar{r}} - \alpha \tilde{T}^{(0)} z_\theta^{(0)}  \\
\label{momentfirstorder_r}  
\left(\frac{v^{(0)}}{\bar{r}}u^{(1)}_\varphi - \frac{2v^{(0)}v^{(1)}}{\bar{r}} + \left(w^{(0)}\right)^2\kappa^{(0)}\cos\varphi\right) \hspace{1.5cm}  \\ \nonumber= - P_{\bar{r}}^{(1)} - \alpha \tilde{T}^{(0)}z_r^{(0)} 
\end{gather} 
\end{subequations}\\
where 
\begin{eqnarray}
z_r^{(0)} &=& z_n^{(0)} \cos\varphi + z_b^{(0)}\sin\varphi \\
z_{\theta}^{(0)} &=&   z_b^{(0)} \cos\varphi - z_n^{(0)}\sin\varphi
\end{eqnarray} \\
For the first order continuity equation we get 
\begin{eqnarray}
\label{contfirstorder}
\sigma^{(0)}\left(v_\varphi^{(1)} + \left(\bar{r}u^{(1)}\right)_{\bar{r}}\right)+\bar{r}w_s^{(0)} \hspace{2cm}\\ \nonumber + \bar{r}\kappa^{(0)}\sigma^{(0)}\sin\varphi\; v^{(0)} = 0 
\end{eqnarray}
while the first order temperature equations is 
\begin{equation}
\label{firstordertemperature}
\frac{v^{(0)}}{\bar{r}}T_\varphi^{(1)} + u^{(1)}T_{\bar{r}}^{(0)} + \frac{w^{(0)}}{\sigma^{(0)} }T_s^{(0)}  = 0
\end{equation}
Note that the right hand side of (\ref{temperatureeqn}) is $\mathcal{O}(1)$ and thus has no impact on the first two order temperature equations. Similarly, the torsion term appearing in (\ref{IncomDensityeq}) is $\mathcal{O}(\epsilon)$ and thus only enters in the 2nd order equations.  
\\\\ For the inner solution, we additionally have
\begin{eqnarray}
\label{bc_at_r=0}
    v^{(i)} = 0, \quad u^{(i)} = 0, \quad\quad i=0,1,2,...  \quad \text{  at  } \bar{r}=0
\end{eqnarray}
which comes directly from the definition of the relative velocity $\mathbf{V}$.
\subsection{The evolution equations for the filament}
To leading order, the outer flow equation is axially symmetric (i.e. independent of $s$). Consequently, it would make sense to 
make a similar requirement for the inner solution, as was indeed done by \citet{CallegariTing1978} and \citet{Harikrishnan2023}. In particular this would mean making the assumption that $v_s^{(0)}=0$. However, by averaging  (\ref{momentfirstorder_s}) with respect to $\varphi$ we get
\begin{equation}
\label{thetaaverfirstordermomeq_s}
   u_c^{(1)}w^{(0)}_{\bar{r}} + \frac{w^{(0)}w^{(0)}_s}{\sigma^{(0)}} = -\frac{P_s^{(0)}}{\sigma^{(0)}} -\alpha \tilde{T}^{(0)}z_\text{t}^{(0)}  
\end{equation}
where 
\begin{equation}
\label{symmetric_part_of_f}
f_c = \frac{1}{2\pi}\int_0^{2\pi} f(\varphi) \; d\varphi
\end{equation}
denotes the average with respect to $\varphi$, or equivalently the symmetric part, of the function $f$.  Applying the same averaging procedure to  (\ref{momentfirstorder_theta}) and (\ref{contfirstorder}) yields
\begin{eqnarray}
\label{thetaaverfirstordermomeq_theta}
    \frac{1}{\bar{r}}\left(\bar{r}v^{(0)}\right)_{\bar{r}}\;u_c^{(1)} + \frac{w^{(0)}v_s^{(0)}}{\sigma^{(0)}}= 0 \\
    \sigma^{(0)}\left(\bar{r}u_c^{(1)}\right)_{\bar{r}} + \bar{r}w_s^{(0)} = 0
\end{eqnarray}
Hence, if $v_s^{(0)}=0$, we must have that 
\begin{equation}
u^{(1)}_c = 0 \quad \text{ and } \quad  w^{(0)}_s = 0
\end{equation}
but we cannot from this conclude that $P_s^{(0)}=0$, due to the buoyancy term appearing in (\ref{thetaaverfirstordermomeq_s}). Unless the buoyancy term vanishes, this leads to a contradiction as \ref{momentleadingorder_r}) relates $P^{(0)}$ directly to $v^{(0)}$. Thus, we necessarily must require that the leading order components are all axially dependent. The case of vortex filaments with leading order axial dependence was treated by \citet{KleinTing1992}, again excluding the force of gravity. {We extend here their one-time scale analysis to include gravity, and do not address the associated two-time-scale analysis in this paper. For a more complete discussion of axisymmetric axial variations on vortices, and two-time scale analysis, the reader is referred to e.g. \citet{Childress2021}, \citet{TingKleinKnio2007} (section 3.3.3 and 3.3.7) and \citet{Margerit2002}}. \\\\
Since the first order equations are linear with respect to the first order field components, $u^{(1)}, v^{(1)}, w^{(1)}, \tilde{T}^{(1)}$ and $P^{(1)}$, we can decompose the solution into symmetric and asymmetric parts, i.e. \[f(\varphi) = f_c + f_a(\varphi)\] with $f_c$ given by (\ref{symmetric_part_of_f}). It is interesting to to note that when splitting (34) between its symmetric and asymmetric parts, the added buoyancy term in (\ref{momentfirstorder_s}) will impact only the symmetric part of the equations. On the contrary, the added buoyancy term in (\ref{momentfirstorder_theta}) and (\ref{momentfirstorder_r}) will impact only the asymmetric part of the equations. The asymmetric first order momentum equations are thus:
\begin{subequations}
\begin{gather} \label{assymetric_1st_order_momentum_eqn_w}
    u_a^{(1)}w_{\bar{r}}^{(0)} + v^{(0)}\frac{\left(w_a^{(1)}\right)_\varphi}{\bar{r}} + w^{(0)}\kappa^{(0)}v^{(0)}\sin\varphi = 0 \qquad\qquad \\
\label{assymetric_1st_order_momentum_eqn_v}
    u_a^{(1)}v_{\bar{r}}^{(0)} + \frac{v^{(0)}\left(v^{(1)}_a\right)_{\varphi}}{\bar{r}} + \frac{v^{(0)}u^{(1)}_a}{\bar{r}} - \left(w^{(0)}\right)^2\kappa^{(0)}\sin\varphi  \\ \nonumber = - \frac{\left(P^{(1)}_a\right)_\varphi}{\bar{r}} - \alpha \tilde{T}^{(0)} z_{\theta}^{(0)}\qquad\qquad \\
\label{assymetric_1st_order_momentum_eqn_u}
    \frac{v^{(0)}}{\bar{r}}\left(u_a^{(1)}\right)_\varphi-\frac{2v^{(0)}v_a^{(1)}}{\bar{r}} + \left(w^{(0)}\right)^2\kappa^{(0)}\cos\varphi  \\ \nonumber = -\left(P^{(1)}_a\right)_{\bar{r}} - \alpha \tilde{T}^{(0)} z_{r}^{(0)} \qquad\qquad
\end{gather}
\end{subequations}
Hence, we conclude that since the $s$-derivatives appear exclusively in the symmetric terms, these terms do not appear in the asymmetric first order equations, from which we will derive the filament equation of motion and the associated core constant. 
As a final remark on the topic of the gravity induced axial asymmetry, we note that if the vortex filament is placed horizontally such that the term proportional to $z_\text{t}^{(0)}$ vanishes, one would again be permitted to assume axial independence of the leading order velocity components. Thus, it is only for filaments that are tilted with respect to the vertical for which the axial dependence necessarily enters.  
\\\\
We continue by first noting that (\ref{assymetric_1st_order_momentum_eqn_w}) decouples from the other first order momentum equations, by which we mean that first order tangential velocity component, $w^{(1)}$ only appears in this equation and not in the other two.  Furthermore, we can eliminate $P_c^{(1)}$ from (\ref{assymetric_1st_order_momentum_eqn_v}) and (\ref{assymetric_1st_order_momentum_eqn_u}) through cross differentiation. Through introduction of the asymmetric stream function $\psi^{(1)}$ 
\begin{equation}
\label{streamfunction}
    u_a^{(1)} = \frac{1}{\bar{r}}\psi_\varphi^{(1)} \quad ,\quad v_a^{(1)} = - \psi^{(1)}_{\bar{r}} + \bar{r}\kappa^{(0)}v^{(0)}\cos\varphi
\end{equation}
and the leading order axial vorticity, $\zeta^{(0)}$
\begin{equation}
\label{leading_order_circulation}
    \zeta^{(0)} = \frac{1}{\bar{r}}\left(\bar{r}v^{(0)}\right)_{\bar{r}}
\end{equation}
the two first order momentum equations, (\ref{assymetric_1st_order_momentum_eqn_v}) and (\ref{assymetric_1st_order_momentum_eqn_u}), are transformed into one differential equation for the stream function $\psi^{(1)}$:
\begin{eqnarray}
    \label{psi_pde}
    v^{(0)}\Delta\psi^{(1)}_\varphi -\zeta_{\bar{r}}^{(0)}\psi^{(1)}_{\varphi} = &-&\kappa^{(0)}\sin\varphi\; H(\bar{r},s,t)\nonumber \\ 
     &+& \alpha B(\bar{r},s,t)z_\theta^{(0)} 
\end{eqnarray}
where $\Delta$ denotes the Laplacian of $\bar{r}$ and $\varphi$ {defined as 
$$   
\Delta = \frac{1}{\bar{r}}\frac{\partial}{\partial \bar{r}} \left(\bar{r}\frac{\partial}{\partial \bar{r}}\right) + \frac{1}{\bar{r}^2}\frac{\partial^2}{\partial \varphi^2}
$$
}
and
\begin{eqnarray}
\label{H}
    H(\bar{r},s,t) &=& 2\bar{r}\zeta^{(0)} v^{(0)} + \left(v^{(0)}\right)^2 + 2\bar{r}w^{(0)}w^{(0)}_{\bar{r}} \\
\label{B}
    B(\bar{r},s,t) &=& \bar{r}\tilde{T}^{(0)}_{\bar{r}}
\end{eqnarray}
We expand $\psi^{(1)}$ in a Fourier series as follows
\begin{equation}
\label{FourierExpansion}
    \psi^{(1)} = \sum_{n=1}^{\infty} \left(\psi_{n1}^{(1)}\cos n\varphi + \psi_{n2}^{(1)}\sin n\varphi\right)
\end{equation}
By using (\ref{FourierExpansion}) in  (\ref{psi_pde}) and identifying all  Fourier modes in the equation, we get an equation for each mode. This set of equations can be written in the following compact form
\begin{eqnarray}
\label{asym_psi_ODE}
v^{(0)}\left[\frac{1}{\bar{r}}\frac{\partial}{\partial \bar{r}} + \frac{\partial^2}{\partial \bar{r}^2} - \left(\frac{n^2}{\bar{r}^2} + \frac{\zeta_{\bar{r}}^{(0)}}{v^{(0)}}\right)\right]\psi^{(1)}_{nj} = \Upsilon^{(0)}_{nj}
\end{eqnarray}
where
\begin{equation}
\label{Upsilon}
\Upsilon^{(0)}_{nj} = \left[\kappa^{(0)} H + \alpha z_\text{n}^{(0)}  B \right]\delta_{n1}\delta_{j1}   
+ \alpha z_\text{b}^{(0)} B \delta_{n1}\delta_{j2}
\end{equation} 
The boundary conditions for (\ref{asym_psi_ODE}) are given by (\ref{bc_at_r=0}) and the farfield conditions. The latter are found by comparing the inner expansion of the circumferential and tangential velocity components,  (\ref{expansion_v}) and (\ref{expansion_w}), with the outer expansion,  (\ref{outer_solution}). For the leading order velocity components we get\\
\begin{subequations}
\begin{eqnarray}
\label{boundcond_v0}
v^{(0)} &\sim& \frac{{1}}{2\pi \bar{r}} \quad \text{ as }\bar{r}\longrightarrow\infty \\
\label{boundcond_w0}
    w^{(0)} &\sim& o(\bar{r}^{-n}) \quad \text{ for all } n \text{ as } \bar{r}\longrightarrow\infty 
\end{eqnarray}
\end{subequations}\\
while for the first order components we get
\begin{subequations}
\begin{eqnarray}
\label{boundcond_v1}
v^{(1)} &\longrightarrow& \frac{ \kappa^{(0)}}{4\pi} \cos\varphi \left(\ln\frac{1}{\epsilon\bar{r}} + 1\right) \\ \quad\quad\quad &\quad&\nonumber + \left(\mathbf{Q}_0-\dot{\mathbf{X}}^{(0)}\right)\cdot\hat{ \theta}^{(0)}  \\
\label{boundcond_u1}
u^{(1)} &\longrightarrow& \frac{ \kappa^{(0)}}{4\pi } \sin \varphi \ln\frac{1}{\epsilon\bar{r}}  + \left(\mathbf{Q}_0-\dot{\mathbf{X}}^{(0)}\right)\cdot\hat{\mathbf{r}}^{(0)}  \hspace{1cm}\\
\label{boundcond_w1}
w^{(1)} &\longrightarrow& \mathbf{Q}_0\cdot\hat{\tau}^{(0)} 
\end{eqnarray}\\
\end{subequations}
as $\bar{r}\longrightarrow\infty$. Here $\mathbf{Q}_0  = \left(\mathbf{Q}_f+\mathbf{Q}_2(r=0)\right)$. {Note that we have written the matching conditions, (\ref{boundcond_v0})-(\ref{boundcond_w1}), in dimensionless form, which is achieved by dividing the dimensional equations by $\Gamma/L$.} 
Given this, (\ref{asym_psi_ODE}) implies that
\[
\psi_{n1}^{(1)} = 0 \quad,\quad \psi_{n2}^{(1)} = 0 \quad\text{for } n\neq 1
\]\\
The inhomogeneous solutions to (\ref{asym_psi_ODE}), for $n=1$, that satisfies the boundary condition at $\bar{r}=0$  is given by
\begin{equation}
\label{solution_upsilon}
    \psi_{1j}^{(1)} = v^{(0)} \int_0^{\bar{r}} \frac{1}{z(v^{(0)})^2}\left[\int_0^{z}\xi \Upsilon^{(0)}_{nj}(\xi,s,t)\; d\xi\right]dz
\end{equation}
To confirm this assertion, observe that both $v^{(0)}$ and $v^{(0)}\int_0^{\bar{r}}\frac{1}{z\left(v^{(0)}\right)^2}\;dz$ are solutions to the homogeneous version of  (\ref{asym_psi_ODE}). 
This provides the  $\psi^{(1)}_{1j}$ functions
\begin{subequations}
\begin{equation}
\label{solution_psi}
    \psi^{(1)}_{11} = \kappa^{(0)} v^{(0)} J + \alpha z_\text{n}^{(0)}  v^{(0)} I
\end{equation}
and
\begin{equation}
\label{solution_psi12}
    \psi^{(1)}_{12} = \alpha z_\text{b}^{(0)}  v^{(0)} I
\end{equation}
\end{subequations}
where 
\begin{subequations}
\begin{equation}
\label{J}
J =   \int_0^{\bar{r}} \frac{1}{z(v^{(0)})^2}\left[\int_0^{z}\xi H(\xi,s,t)\; d\xi\right]dz
\end{equation}
and
\begin{equation}
\label{I}
 I =  \int_0^{\bar{r}} \frac{1}{z(v^{(0)})^2}\left[\int_0^{z}{\xi}^2\tilde{T}^{(0)}_{\xi}\; d\xi\right]dz
\end{equation}
\end{subequations}
as $B(\xi,s,t)=\xi\tilde{T}^{(0)}_\xi$.
\\\\
It remains to determine the correct scaling for the leading order temperature difference $\tilde{T}^{(0)}$. The temperature satisfies the parabolic-transport equation. Thus, since $\tilde{T}^{(0)}$ is concentrated on the vortex initially, we can conclude that the temperature should decay exponentially as $\bar{r}$ goes to infinity for all later time $t$. In other words, 
\begin{equation}
\label{boundcond_T0}
    \tilde{T}^{(0)} \sim o(\bar{r}^{-n}) \quad \text{for all } n \text{ as } \bar{r} \longrightarrow \infty 
\end{equation}
For completeness we note that in the more realistic case of non-zero Mach number, the temperature also depends on the pressure, which would yield a different scaling for the leading order temperature difference; see \citet{TingKleinKnio2007} for a description of fully compressible vortex flows, excluding the force of gravity. 
\\\\
Inserting (\ref{boundcond_v0}), (\ref{boundcond_w0}) and (\ref{boundcond_T0}) into (\ref{Upsilon}) yields, 
\begin{equation}
\Upsilon^{(0)}_{nj} \sim   \frac{\kappa^{(0)}}{2\pi \bar{r}} 
\end{equation}
as $\bar{r}\longrightarrow\infty$. 
By integrating (\ref{asym_psi_ODE}) for large $\bar{r}$ twice, we get  
\begin{subequations}
\begin{eqnarray}
\label{solution_psi_asymptotic}
    \psi_{11}^{(1)} &\sim& \frac{\kappa^{(0)}}{4\pi}\bar{r} \ln \bar{r} + C_1 \bar{r} \\
    \label{solution_asymptotic_psi12}
    \psi_{12}^{(1)} &\sim& C_2 \bar{r}
\end{eqnarray}
\end{subequations}
where $C_1$  and $C_2$ are integration constants which together are referred to as the \textit{core constant}. As will be demonstrated shortly, $C_1$ and $C_2$ correspond to the binormal and normal parts of the core constant and couple the velocity of the filament to the motion inside the filament core. \\\\
Comparing (\ref{solution_psi}) and  (\ref{solution_psi_asymptotic}) we find that 
\begin{eqnarray}
    C_1(s,t) &=&  \lim_{\bar{r}\longrightarrow \infty} (\psi^{(1)}_{11}- \frac{\kappa^{(0)}}{4\pi}\bar{r}\ln\bar{r})/\bar{r} \\ 
    &=& \frac{{1}}{4\pi}\left[\kappa^{(0)}\lim_{\bar{r}\longrightarrow \infty} ( J -h\ln\bar{r})/h 
    + \alpha z_{\text{n}}^{(0)}  \lim_{\bar{r}\longrightarrow \infty} I/h\right] \nonumber
\end{eqnarray}
where $h = \frac{{1}}{4\pi}\bar{r} /v^{(0)}$, and $h(\infty) = \bar{r}^2/2 $. {Similarly, by comparing (\ref{solution_psi12})  and (\ref{solution_asymptotic_psi12}) we get}
\begin{eqnarray}
C_2(s,t) =  \lim_{\bar{r}\longrightarrow \infty} (\psi^{(1)}_{12}/\bar{r}) 
         = \frac{{1}}{4\pi} \alpha z_\text{b}^{(0)}  \lim_{\bar{r}\longrightarrow \infty} I/h 
\end{eqnarray}
The Hopital's rule is applied to those limits:
having $f(x)$ and $g(x)$ that goes to infinity at $x=\infty$, if $f_x/g_x(\infty)=l$ then $f/g(\infty)=l$. Hence, we get
\begin{eqnarray}
    \lim_{\bar{r}\longrightarrow \infty} ( J -h\ln\bar{r})/h  = C_v(s,t) + C_w(s,t)
\end{eqnarray}
where 
\begin{eqnarray}
	\label{Core_constant}
	C_v(s,t) &=& \lim_{\bar{r}\longrightarrow \infty}\left(4\pi^2\int_0^{\bar{r}}\xi\left(v^{(0)}\right)^2\;d\xi -\ln\bar{r}\right) +\frac{1}{2} 
\\
	\label{Core_constant_TingKlein}
	C_w(s,t) &=& -8\pi^2\int_0^{\infty}\xi\left(w^{(0)}\right)^2\;d\xi
\end{eqnarray}
The details of this calculation are included in Appendix A. {Let us define} 
\[
   C_T(s,t) = \lim_{\bar{r}\longrightarrow \infty} I/{h}
\]
{We get}
\begin{eqnarray}
\label{Core_constant_gravity_part}
   C_T(s,t) 
    &=&  4\pi^2\lim_{\bar{r}\longrightarrow \infty}
    \int_0^{\bar{r}}{\xi}^2\tilde{T}^{(0)}_{\xi}\; d\xi\nonumber\\
    &=&  -8\pi^2
    \int_0^{\infty}\xi\tilde{T}^{(0)}\; d\xi
\end{eqnarray}
{where we have used the asymptotic behaviour of the temperature perturbation, as stated in (\ref{boundcond_T0}), to establish that the integral in (\ref{Core_constant_gravity_part}) converges at infinity.}
And so we get for the core constants $C_1(s,t)$ and $C_2(s,t)$
\begin{subequations}
\begin{eqnarray}
    C_1(s,t) &=& \frac{{1}}{4\pi}\left[\kappa^{(0)}(C_v(t) + C_w(t))
    + \alpha z_\text{n}^{(0)}  C_T(t)\right] 
    \label{Coreconstant1}   \\
C_2(s,t) &=& \frac{{1}}{4\pi} \alpha z_\text{b}^{(0)}  C_T(s,t) 
\label{coreconstant2}
\end{eqnarray}
\end{subequations}
To arrive at an equation of motion for the filament, we need a matching condition for $\psi^{(1)}$ to which  (\ref{solution_psi_asymptotic}) and (\ref{solution_asymptotic_psi12})  can be compared. This condition is given by the matching conditions for the leading order velocity field components,  (\ref{boundcond_v1})-(\ref{boundcond_w1}).
We get
\begin{gather}\label{solution_psi1_asym}
    \psi^{(1)} \sim - \frac{\kappa^{(0)}}{4\pi} \cos\varphi\;\bar{r}\ln\left(\frac{1}{\epsilon \bar{r}}\right) \hspace{2cm}\\ \nonumber + \bar{r}\left[(\mathbf{Q}_0-\dot{\mathbf{X}}^{(0)})\cdot\left(\hat{\mathbf{n}}^{(0)}\sin\varphi - \hat{\mathbf{b}}^{(0)}\cos\varphi\right)\right]
\end{gather}
Comparing (\ref{solution_psi_asymptotic}), (\ref{solution_asymptotic_psi12}) and (\ref{solution_psi1_asym}), using  (\ref{FourierExpansion}), yields $C_1(s,t)$ and $C_2(s,t)$ in terms of the filament velocity, from which we write
\begin{equation}
\label{filament_velocity_all_components}
    \dot{\mathbf{X}}^{(0)} =  \mathbf{Q}_0^{\perp} + \left(\frac{\kappa^{(0)}}{4\pi}\ln \frac{1}{\epsilon}  + C_1(s,t)\right)\hat{\mathbf{b}}^{(0)} - C_2(s,t) \hat{\mathbf{n}}^{(0)}
\end{equation}
where $\mathbf{Q}_0^{\perp} = \mathbf{Q}_0 - \mathbf{Q}_0 \cdot \hat{\mathbf{\tau}}^{(0)} \hat{\mathbf{\tau}}^{(0)}$ denotes the part of $\mathbf{Q}_0$ which is perpendicular to $\hat{\tau}$. Then inserting the previously found expressions for  $C_1(s,t)$ and $C_2(s,t)$, given by (\ref{Coreconstant1}) and (\ref{coreconstant2}), into (\ref{filament_velocity_all_components}) gives the equation of motion 
\begin{eqnarray}
\label{buoyancy_eq_of_motion}
 \dot{\mathbf{X}}^{(0)} =  \mathbf{Q}_0^{\perp} &+&\frac{\kappa^{(0)}}{4\pi}\left[\ln \frac{1}{\epsilon} + C_v(s,t) + C_w(s,t)\right]\hat{\mathbf{b}}^{(0)}\nonumber \\  
 &-& \frac{{1}}{4\pi}  \alpha C_T(s,t) \hat{\mathbf{z}}\times \hat{\mathbf{\tau}}^{(0)} 
\end{eqnarray}
where we used  $\hat{\mathbf{z}}\times \hat{\mathbf{\tau}}^{(0)}  = \hat{\mathbf{z}} \times (\hat{\mathbf{n}}^{(0)} \times \hat{\mathbf{b}}^{(0)} ) = \hat{\mathbf{n}}^{(0)} z_\text{b}^{(0)}  -\hat{\mathbf{b}}^{(0)} z_\text{n}^{(0)} $. To close the system, we need equations describing the temporal and spatial evolution of $w^{(0)}$, $v^{(0)}$, $P^{(0)}$ and $T^{(0)}$; these equations will be derived in the next section.\\\\
\begin{figure}
    \centering
    {\frame{\includegraphics[width = 80mm]{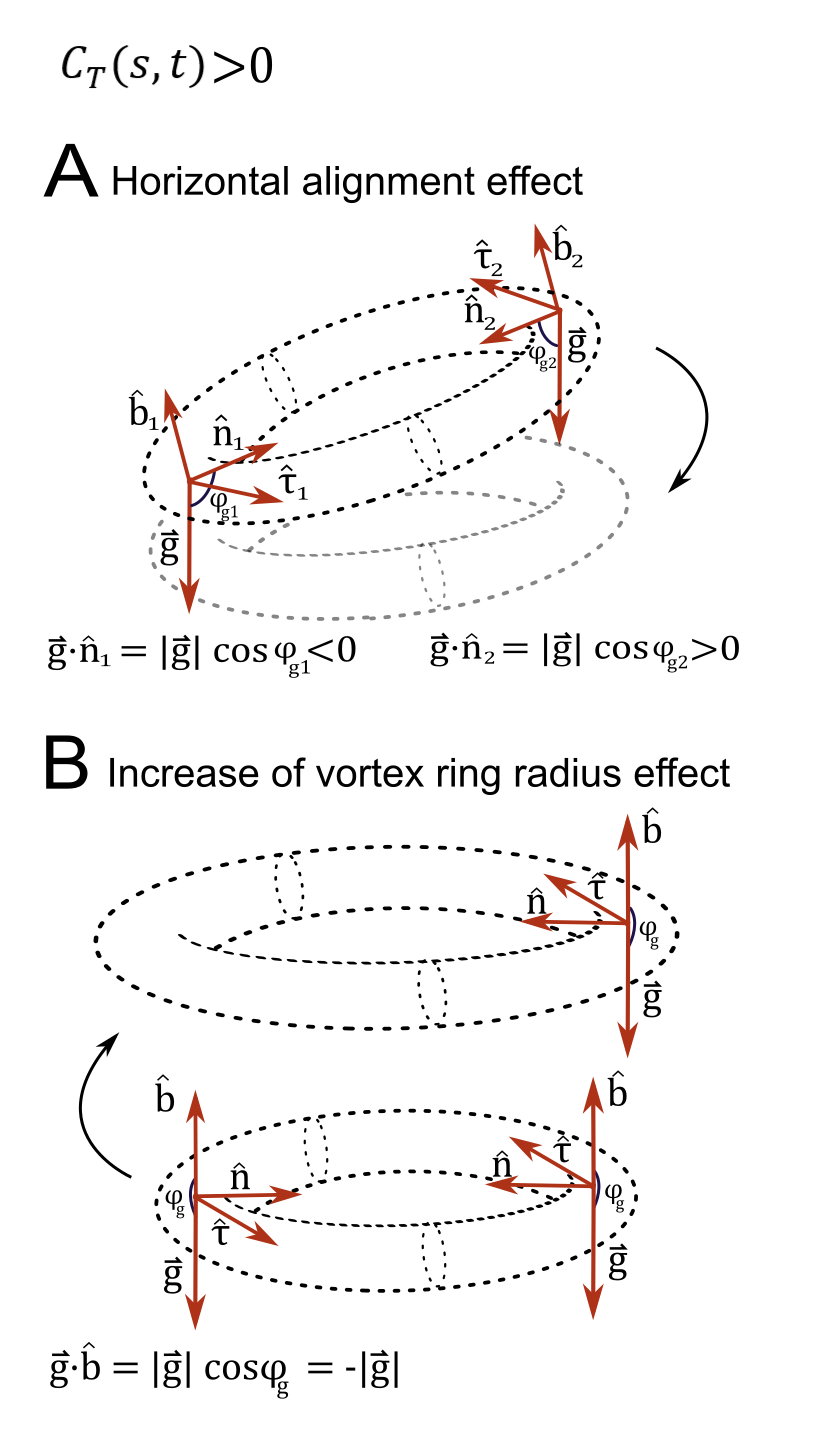}}
    \caption{An illustration of the impact of gravity on a vortex ring assuming that $C_T(s,t) > 0$ for all $s$ and $t$. This would e.g. be the case if the density of the filament is everywhere lower than the density of the surrounding fluid such that $\Tilde{T}^{(0)} < 0$ for all $s$, $t$ and $\bar{r}$. In this simple example, gravity acts both to align the vortex ring with the horizontal plane (A), and causes the ring to expand as it rises (B).}}
    \label{fig:VortexSketch}
\end{figure}
We conclude this section with a discussion of the implications of the inclusion of buoyancy for the filament equation of motion. Firstly, we note that the binormal part of the buoyancy term appearing in (\ref{buoyancy_eq_of_motion}) is maximal for a horizontally aligned filament, and vanishes for a filament aligned in the vertical direction. 
Assuming the density of the core is smaller than the density of the surrounding fluid, such that $\tilde{T}^{(0)}$ is negative and $C_T$ positive, then this term is negative whenever the angle between the normal and vertical unit vectors, $\hat{\mathbf{n}}$ and $\hat{\mathbf{z}}$, is between $0$ and $\pi/2$, and positive whenever the angle is between $\pi/2$ and $\pi$. Thus, this term is responsible for the tendency of buoyant vortex rings to align themselves in the horizontal plane, as noted by \citet{ChangSmith2018}. {A schematic illustration of this phenomenon is shown in Figure 2 (A).} 
Secondly, for a circular, horizontally aligned filament, {as shown in Figure 2 (B)}, the normal part of the buoyancy term appearing in (\ref{buoyancy_eq_of_motion}) yields, provided $\tilde{T}^{(0)}$ is negative, a negative contribution to the normal component of the filament velocity. This, in turn will cause the ring to expand, with the rate of expansion being determined by the density difference between the filament and the surrounding fluid.
Next, for a very long vertically aligned filament we would expect there to be a considerable density difference between the top and the bottom of the filament. If such a filament were to be slightly tilted in such a way that either $z_{\text{b}}^{(0)} $ or $z_{\text{n}}^{(0)} $ are no longer identically zero, one would see a potentially significant difference in the normal or binormal velocity component arising from the core constant, depending on the height above ground. In that case, the lower, and presumed denser, part of the filament would have a smaller velocity than the parts further up, causing the filament to tilt further. However, this effect would be competing with the effect of the tangential leading order velocity, which has been found to reduce the velocity in the binormal direction (\citet{CallegariTing1978}), an effect that, as we shall see in the next section, is potentially enhanced by the presence of gravity, depending on the direction of the axial flow. Furthermore, we note that if one were to employ the anelastic approximation and thus properly account for the stratification of the external flow field, it is conceivable that the difference in $\tilde{T}^{(0)}$ between the lower and upper parts of the filament would be reduced as the background density is no longer a constant. 
Finally, we would like to highlight the similarity between the gravitational terms as they appear in (\ref{Coreconstant1}) and (\ref{coreconstant2}), and (2.24) in \citet{ChangSmith2018}. The only difference is that in the latter case the density within the core is assumed to be constant, while we allow the density of the core to continuously approach the density of the surrounding fluid as one approaches the outer boundary. Thus, we have succeeded in reproducing the results of \citet{ChangSmith2018} using a different approach to estimate the velocity of the vortical core. It should be noted that we have limited our analysis to finding the leading order contribution to the filament velocity, i.e. $\dot{\mathbf{X}}^{(0)}$, and one could in principle continue the analysis to first order following the method outlined in \citet{Fukumoto1991}and \citet{Margerit1996}. This should lead to terms corresponding to the last three terms in (2.24) in  \citet{ChangSmith2018}. For completeness, we note that one could derive  (\ref{Coreconstant1}) and (\ref{coreconstant2}) directly from the Biot-Savart law and vorticity equation, following the procedure outlined in \citet{KleinKnio1995}.

\section{Evolution Equations for the Leading Order Velocity Components and Temperature}
\label{sec:compconditions}
The evolution equations for the dynamic variables $v^{(0)}$, $w^{(0)}$, $P^{(0)}$ and $T^{(0)}$ are derived from the first and second order \textit{symmetric} density, momentum and temperature equations. As already mentioned, due to the gravity term appearing (\ref{momentfirstorder_s}), we cannot assume that all of the leading order field components are independent of the axial coordinate $s$. However, we could, thus far, seemingly ignore this issue, since at first order the $s$-derivative terms only appear in the symmetric terms, whereas the velocity depends on the first order Fourier modes (see \citet{KleinTing1992}). Since the compatibility conditions, from which we get the evolution equations for the leading order field components,  are given by the symmetric equations, we will now see the appearance of terms containing leading order axial derivatives. For this reason we cannot derive the compatibility conditions the way outlined in  \citet{CallegariTing1978}, but will instead follow the procedure outlined in  \citet{KleinTing1992}, in which the dynamic variables were {assumed} to have leading order axial variation. The key observation is that along stream lines in the $r$-$s$-plane the leading order circulation and leading order total head (as will be defined in Section IV.B) are conserved quantities. It should be noted, however, that in defining the s-derivate along a stream line the assumption is made that the leading order axial velocity $w^{(0)}$ does not vanish at any point. A discussion of the case of vanishing axial velocity is left to Appendix C. \\
In broad strokes, the derivation of the evolution equations is presented as follows: first we provide the relevant leading and first order symmetric temperature and momentum equations. Next, the definition of the new dynamic variables, namely circulation and total head, are given and the kinetic energy equation, derived by contracting (\ref{NS_equation_Boussniesq_approx}) with $(\dot{\mathbf{X}}+ \mathbf{V})$, is introduced, before we present the coordinate transformation which defines the s-derivative along a stream line and the associated inverse operator or integral. Finally, by combining the first and second order symmetric equations we derive temporal evolution equations for the dynamic variables $v^{(0)}$, $w^{(0)}$, $P^{(0)}$ and $T^{(0)}$, with the assumption that the filament remains symmetric in some stream-wise direction.   
\\\\
In preparation for this derivation, we recall that we can write the leading order field components as follows:
\begin{equation}
    f^{(1)} = f_c^{(1)} + f_{11}^{(1)} \cos{\varphi} + f_{12}^{(1)}\sin\varphi
\end{equation}
Here $f$ stand for the variables $ u, w, v$ or $T$, and $f_{11}$ and $f_{12}$ are the Fourier coefficients, which are known functions of $\psi^{(1)}_{11}$ and $\psi^{(1)}_{12}$. From (\ref{streamfunction}) we immediately have that
\begin{eqnarray}
\label{u_11_fourier_component}
    u_{11}^{(1)}  &=& \frac{1}{\bar{r}} \psi_{12}^{(1)} \\
    u_{12}^{(1)}  &=& -\frac{1}{{\bar{r}}}\psi_{11}^{(1)}\\
    v_{11}^{(1)}  &=& -\left(\psi_{11}^{(1)}\right)_{\bar{r}} +\bar{r}\kappa^{(0)}v^{(0)} \\
    v_{12}^{(1)}  &=& -\left(\psi^{(1)}_{12}\right)_{\bar{r}}
\label{v_12_fourier_component}
\end{eqnarray}
while by combining (\ref{streamfunction}), (\ref{firstordertemperature}) and (\ref{assymetric_1st_order_momentum_eqn_w}), we get
\begin{eqnarray}
\label{T_11_fourier_component}
    T_{11}^{(1)} &=& - \psi_{11}^{(1)} \frac{T_{\bar{r}}^{(0)}}{v^{(0)}} \\
    T_{12}^{(1)} &=& - \psi_{12}^{(1)} \frac{T_{\bar{r}}^{(0)}}{v^{(0)}} \\    
    w_{12}^{(1)} &=& - \frac{\psi_{12}^{(1)}w_{\bar{r}}^{(0)}}{v^{(0)}} \\
    w_{11}^{(1)} &=&  - \frac{\psi_{11}^{(1)}w_{\bar{r}}^{(0)}}{v^{(0)}} + \bar{r}w^{(0)} \kappa^{(0)}
\label{w_11_fourier_component}
\end{eqnarray} 
These relations are needed when deriving the symmetric second order momentum and temperature equations.
\subsection{The first and second order symmetric momentum and temperature equations}
The symmetric first order continuity and temperature equations are
\begin{subequations}
\begin{eqnarray}
\label{symmetric_first_order_continutiy}
    \sigma^{(0)} \left(\bar{r}u_c^{(1)}\right)_{\bar{r}} + \bar{r}w_s^{(0)} &=& 0 \\
\label{symetric_first_order_temperature}
    u_c^{(1)} T_{\bar{r}}^{(0)} + \frac{w^{(0)}}{\sigma^{(0)}}T_s^{(0)} &=& 0
\end{eqnarray}
\end{subequations}
while the symmetric first order momentum equations are given by
\begin{subequations}
\begin{eqnarray}
\label{symmetric_first_order_moment_eq_s}
    u_c^{(1)}w_{\bar{r}}^{(0)} + \frac{w^{(0)}w_s^{(0)}}{\sigma^{(0)}} = - \frac{P_s^{(0)}}{\sigma^{(0)}}-\alpha\tilde{T}^{(0)}z_{\text{t}}^{(0)} \\
\label{symmetric_first_order_moment_eq_theta}
    u_c^{(1)} v^{(0)}_{\bar{r}} + \frac{v^{(0)}u_c^{(1)}}{\bar{r}} + \frac{w^{(0)}}{\sigma^{(0)}}v_s^{(0)} = 0 \\
\label{symmetric_first_order_moment_eq_r}
    \frac{2v^{(0)}v_c^{(1)}}{\bar{r}} = \left(P^{(1)}_c\right)_{\bar{r}}
\end{eqnarray}
\end{subequations}
The symmetric part of the second order circumferential momentum equation is: 
\begin{gather}  
    v_t^{(0)} + u_c^{(1)}\left(v^{(1)}_c\right)_{\bar{r}} + \frac{v^{(0)}}{\bar{r}}u_c^{(2)} +u_c^{(2)}v^{(0)}_{\bar{r}} + \frac{v_c^{(1)}u_c^{(1)}}{\bar{r}} \\+ \nonumber \frac{w^{(0)}}{\sigma^{(0)}}\left(v_c^{(1)}\right)_s + \frac{w_c^{(1)}v_s^{(0)}}{\sigma^{(0)}} - \frac{w^{(0)}v_s^{(0)}\sigma^{(1)}}{\left(\sigma^{(0)}\right)^2}  - \kappa^{(0)}w^{(0)}w_{12}^{(1)}
    \\ \nonumber  
    +\frac{1}{2}u_{11}^{(1)} \left(v_{11}^{(1)}\right)_{\bar{r}} +\frac{1}{2}u_{12}^{(1)}\left(v_{12}^{(1)}\right)_{\bar{r}} + \frac{v_{11}^{(1)}u_{11}^{(1)}}{2\bar{r}} + \frac{v_{12}^{(1)}u_{12}^{(1)}}{2\bar{r}}
    \\ \nonumber
    = \frac{\bar{\nu}}{\bar{r}}\left(\bar{r}v_{\bar{r}}^{(0)}\right)_{\bar{r}} - \frac{\bar{\nu}}{\bar{r}^2}v^{(0)} - \frac{1}{2} \alpha ( \tilde{T}_{11}^{(1)}z_b^{(0)} - \tilde{T}_{12}^{(1)}z_n^{(0)})
\end{gather}
which upon introduction of (\ref{u_11_fourier_component})-(\ref{w_11_fourier_component}), using (\ref{psi_pde}) and (\ref{solution_psi}), simplifies to
\begin{gather}
   v_t^{(0)} + u_c^{(1)}\left(v^{(1)}_c\right)_{\bar{r}} + \frac{v^{(0)}}{\bar{r}}u_c^{(2)} +u_c^{(2)}v^{(0)}_{\bar{r}} + \frac{v_c^{(1)}u_c^{(1)}}{\bar{r}} \nonumber\\+  \frac{w^{(0)}}{\sigma^{(0)}}\left(v_c^{(1)}\right)_s + \frac{w_c^{(1)}v_s^{(0)}}{\sigma^{(0)}} - \frac{w^{(0)}v_s^{(0)}\sigma^{(1)}}{\left(\sigma^{(0)}\right)^2} 
   \label{second_ord_momentum_eq_theta}  \\ \nonumber =  \frac{\bar{\nu}}{\bar{r}}\left(\bar{r}v_{\bar{r}}^{(0)}\right)_{\bar{r}} - \frac{\bar{\nu}}{\bar{r}^2}v^{(0)}  + \frac{\alpha}{2}\kappa^{(0)} z_b^{(0)} v^{(0)} I\zeta^{(0)}
\end{gather}
where $I$ is given by (\ref{I}).\\\\
Next, the symmetric part of the second order tangential momentum equation is:
\begin{gather}  
    w_t^{(0)} + u_c^{(2)}w^{(0)}_{\bar{r}} + u_c^{(1)}\left(w_c^{(1)}\right)_{\bar{r}} + \frac{w^{(0)}}{\sigma^{(0)}}\left(w_c^{(1)}\right)_s + \frac{w^{(1)}_cw_s^{(0)}}{\sigma^{(0)}} \nonumber\\  - \frac{\sigma^{(1)}w^{(0)}w^{(0)}_s}{\left(\sigma^{(0)}\right)^2}    + \hat{ \tau}^{(0)}\cdot\dot{\mathbf{X}}_s^{(0)} \frac{w^{(0)}}{\sigma^{(0)}} 
    = - \frac{1}{\sigma^{(0)}}\left(P_c^{(1)}\right)_s \quad \nonumber \\
     + \frac{\sigma^{(1)}P_s^{(0)}}{\left(\sigma^{(0)}\right)^2} + \frac{\bar{\nu}}{\bar{r}}\left(\bar{r}w_{\bar{r}}^{(0)}\right)_{\bar{r}} - \alpha \tilde{T}_c^{(1)}z_{\text{t}}^{(0)} - \alpha \tilde{T}^{(0)} z_t^{(1)}\\ \nonumber 
   +\frac{w^{(0)}\kappa^{(0)}}{2}\left(u_{11}^{(1)} - v_{12}^{(1)}\right)  - \frac{1}{2\bar{r}}\left( v_{11}^{(1)}w_{12}^{(1)}-v_{12}^{(1)}w_{11}^{(1)}    
   \right) \\ \nonumber - \frac{\kappa^{(0)}v^{(0)}}{2}w_{12}^{(1)}-\frac{1}{2}\left(u_{11}^{(1)} \left(w_{11}^{(1)}\right)_{\bar{r}}+ u_{12}^{(1)}\left(w_{12}^{(1)}\right)_{\bar{r}}\right) 
\end{gather} 
which, when introducing (\ref{u_11_fourier_component})- (\ref{w_11_fourier_component}), using (\ref{solution_psi}), simplifies to 
\begin{gather}
       w_t^{(0)} + u_c^{(2)}w^{(0)}_{\bar{r}} + u_c^{(1)}\left(w_c^{(1)}\right)_{\bar{r}} + \frac{w^{(0)}}{\sigma^{(0)}}\left(w_c^{(1)}\right)_s + \frac{w^{(1)}_cw_s^{(0)}}{\sigma^{(0)}} \nonumber\\ \label{second_ord_momentum_eq_s} - \frac{\sigma^{(1)}w^{(0)}w^{(0)}_s}{\left(\sigma^{(0)}\right)^2}    + \hat{ \tau}^{(0)}\cdot\dot{\mathbf{X}}_s^{(0)} \frac{w^{(0)}}{\sigma^{(0)}} 
    = - \frac{1}{\sigma^{(0)}}\left(P_c^{(1)}\right)_s \quad \\\nonumber  + \frac{\bar{\nu}}{\bar{r}}\left(\bar{r}w_{\bar{r}}^{(0)}\right)_{\bar{r}} - \alpha \tilde{T}_c^{(1)}z_{\text{t}}^{(0)} - \alpha \tilde{T}^{(0)} z_t^{(1)}  \\ \nonumber+ \frac{1}{2}\kappa^{(0)}\alpha z_b^{(0)} v^{(0)} I\; w^{(0)}_{\bar{r}}
\end{gather}
Finally, the symmetric second order temperature equation is:
\begin{gather} 
    T_t^{(0)} + \frac{w^{(0)}}{\sigma^{(0)}}\left(T_c^{(1)}\right)_s + u_c^{(1)}\left(T_c^{(1)}\right)_{\bar{r}} - \frac{\sigma^{(1)}w^{(0)}}{\left(\sigma^{(0)}\right)^2}T_s^{(0)} + u_c^{(2)}T^{(0)}_{\bar{r}}\nonumber \\  \nonumber 
    + \frac{w_c^{(1)}}{\sigma^{(0)}}T_s^{(0)}   = \frac{\bar{\eta}}{\bar{r}}\left(\bar{r}T^{(0)}_{\bar{r}}\right)_{\bar{r}} - \frac{1}{2}\left(u_{11}^{(1)}\left(T_{11}^{(1)}\right)_{\bar{r}} + u_{12}^{(1)}\left(T_{12}^{(1)}\right)_{\bar{r}}\right) \\  -
    \frac{1}{2\bar{r}}\left(v_{11}^{(1)}T_{12}^{(1)} - v_{12}^{(1)} T_{11}^{(1)}\right)
\end{gather}
which simplifies to 
\begin{gather}
        T_t^{(0)} + \frac{w^{(0)}}{\sigma^{(0)}}\left(T_c^{(1)}\right)_s + u_c^{(1)}\left(T_c^{(1)}\right)_{\bar{r}} - \frac{\sigma^{(1)}w^{(0)}}{\left(\sigma^{(0)}\right)^2}T_s^{(0)} + u_c^{(2)}T^{(0)}_{\bar{r}}\nonumber \\  
    + \frac{w_c^{(1)}}{\sigma^{(0)}}T_s^{(0)}   = \frac{\bar{\eta}}{\bar{r}}\left(\bar{r}T^{(0)}_{\bar{r}}\right)_{\bar{r}} + \frac{1}{2}\kappa^{(0)}\alpha z_b^{(0)} v^{(0)}I T_{\bar{r}}^{(0)}
    \label{symmetric_2nd_order_temperature_equation}
\end{gather}

\subsection{Change of variables and coordinate transformation} 
The circulation around the boundary of a disk of radius $\bar{r}$ is given by 
\begin{eqnarray}
    \tilde{\Gamma}(\bar{r}, \varphi, s,t ; \epsilon) &=& 2\pi \int_0^{\bar{r}}\bar{r}\zeta \; d\bar{r} \\
    &=& 2\pi \mathcal{G}(\bar{r}, \varphi, s,t ; \epsilon)
\end{eqnarray}
where we have defined $\mathcal{G}(\bar{r}, \varphi, s,t ; \epsilon) = \bar{r}v$. At leading and first order we have, respectively
\begin{eqnarray}
    \mathcal{G}^{(0)} = \bar{r}v^{(0)} \\
     \mathcal{G}^{(1)} = \bar{r}v^{(1)}    
\end{eqnarray}
Using (\ref{leading_order_circulation}) and (\ref{symmetric_first_order_moment_eq_theta}), equation (\ref{second_ord_momentum_eq_theta}) can be now written in terms of $\mathcal{G}^{(0)}$ and $\mathcal{G}^{(1)}$:
\begin{gather}  
     \sigma^{(0)}u_c^{(1)}\left(\mathcal{G}^{(1)}_c\right)_{\bar{r}} + \left(\sigma^{(1)}u^{(1)}_c + \sigma^{(0)}u_c^{(2)}\right)\mathcal{G}_{\bar{r}}^{(0)}\nonumber \\ + w^{(0)}\left(\mathcal{G}_c^{(1)}\right)_s  + w^{(1)}_c\mathcal{G}_s^{(0)}     \label{circulation_equation1}
     = -\sigma^{(0)}\mathcal{G}_t^{(0)}\\ + \sigma^{(0)} \bar{\nu} \left(\mathcal{G}_{\bar{r}\bar{r}}^{(0)} \nonumber-\frac{\mathcal{G}_{\bar{r}}^{(0)}}{\bar{r}}\right)    + \frac{\sigma^{(0)}\kappa^{(0)}}{2\bar{r}}\alpha z_b^{(0)} I \;\mathcal{G}^{(0)}  \mathcal{G}_{\bar{r}}^{(0)}
\end{gather}
As will be shown in the section IV.C, (\ref{circulation_equation1}) yields the first of the required evolution equations, for the field component $v^{(0)}$. \\\\
{Adding gravity to the total head, $\mathcal{H},$ (see \citet{KleinTing1992} for the definition of the total head excluding gravitational forces) given the Boussinesq approximation yields the following definition:}
\begin{eqnarray}
\label{definition_total_head_boussinesq}
 {\cal{H}}= P+\left(\dot{\mathbf{X}}+\mathbf{V}\right)^2/2 +  \alpha \epsilon^{-2}\tilde{T} \tilde{\mathcal{Z}}(\bar{r},\varphi, s,t;\epsilon)
\end{eqnarray}
where $\tilde{\mathcal{Z}}$ is the height of the point $\boldsymbol{x}(x,y,z,t) = \boldsymbol{x}(r,\varphi,s,t)$ in curvilinear coordinates, defined through 
\begin{eqnarray*}
    (Z - Z_0)(x,y,z) &=& z-z_0  \\&=& \tilde{\mathcal{Z}}(\bar{r},\varphi,s,t;\epsilon) \\ &=&\left[\mathbf{X}(s,t;\epsilon)+\epsilon\bar{r}\hat{\mathbf{r}}-\mathbf{X}(0,t;\epsilon)\right] \cdot \hat{\mathbf{z}}
\end{eqnarray*}
Here $Z - Z_0$ denotes the height of any point $(x,y,z)$ with respect to the point $Z_0(x,y,z) = z_0$ in Cartesian coordinates.   
The symmetric part of $\tilde{\mathcal{Z}}$ is thus
\begin{eqnarray}\tilde{\mathcal{Z}}_c(s,t {;\epsilon}) &=&  \hat{\mathbf{z}} \cdot \left[\mathbf{X}(s,t{; \epsilon})-\mathbf{X}(0,t{;\epsilon})\right]  \\
&=&{\int_{0}^{s}\sigma(s',t{;\epsilon}) z_\text{t}(s',t{;\epsilon}) ds'}
\end{eqnarray} 
which at leading and first order gives
\begin{eqnarray*}
\label{leading_height}
\tilde{\mathcal{Z}}_c^{(0)}(s,t) &=&  \hat{\mathbf{z}} \cdot \left[\mathbf{X}^{(0)}(s,t)-\mathbf{X}^{(0)}(0,t)\right] \qquad\qquad\\ &=&\int_{0}^{s}\sigma^{(0)}(s',t) z_\text{t}^{(0)}(s',t) ds'\qquad\qquad \end{eqnarray*}
\begin{eqnarray*}
\label{first_order_height}
\tilde{\mathcal{Z}}_c^{(1)}(s,t) &=&  \hat{\mathbf{z}} \cdot \left[\mathbf{X}^{(1)}(s,t)-\mathbf{X}^{(1)}(0,t)\right] \\ &=& \int_{0}^{s}\Big(\sigma^{(1)}(s',t) z_\text{t}^{(0)}(s',t) \\ &\quad& \qquad\qquad + \sigma^{(0)}(s',t) z_t^{(1)}(s',t) \Big) ds'
\end{eqnarray*}
At leading and first order, the symmetric total head is then, respectively
\begin{eqnarray}
\label{leading_order_total_head}
{\cal{H}}^{(0)}&=&p^{(0)}+\frac{{v^{(0)}}^2+{w^{(0)}}^2}{2} +  \alpha {\tilde{T}}^{(0)}\tilde{\mathcal{Z}}_c^{(0)}\\
{\cal{H}}^{(1)}_c&=&p^{(1)}_c +\left[v^{(0)} v^{(1)}_c+ w^{(0)} w^{(1)}_c\right] \qquad \qquad \nonumber \\ \qquad &\quad&+ \alpha \left[{\tilde{T}}^{(1)}_c\tilde{\mathcal{Z}}_c^{(0)}+ \tilde{T}^{(0)}_c \tilde{\mathcal{Z}}_c^{(1)}\right].
\label{first_order_total_head}
\end{eqnarray}
While not explicitly written in (\ref{leading_order_total_head}) and (\ref{first_order_total_head}), the reader should take note that while $\mathcal{H}_c^{(i)}, p_c^{(i)}, v_c^{(i)}, w_c^{(i)}$ and $  \tilde{T}_c^{(i)}$ are functions of $(\bar{r}, s,t)$, $\tilde{\mathcal{Z}}^{(i)}_c$ only depends on $(s,t)$. \\\\
Thus, we see that the effect of the axial component of the gravitational force, which does not enter in the core constant derived in the previous section, is exclusively to alter the total head, or equivalently the kinetic energy, in the moving frame. {For a discussion on the Bernoulli equation, and by extension the total head, in the context of concentrated vortices, the reader is referred to e.g. \citet{Saffmann1993} (section 1.9), \citet{Alekseenko2007} (section 1.2.3) and \citet{TingKleinKnio2007} (sections 3.3.7 and 3.4.5)}. 
\\\\
By contracting (\ref{NS_equation_Boussniesq_approx}) with $\mathbf{\dot{X} + V}$ we get an equation for the kinetic energy, which in terms of the total head can be written as:
\begin{gather}
\label{kinetic_energy_equation}
   \mathcal{H}_t + \mathbf{v}_R\cdot \nabla \mathcal{H} =  P_t - \mathbf{v}_E\cdot\nabla P  + \epsilon^{-2}\alpha \tilde{T} \mathcal{Z}_t   \qquad\\ + \epsilon^2 \bar{\nu} \left(\dot{\mathbf{X} }+\mathbf{V}\right)\cdot \Delta\left(\dot{\mathbf{X} }+\mathbf{V}\right)  - \epsilon^{-2}\alpha \tilde{T}\mathbf{v}_E\cdot\nabla\mathcal{Z} +  \alpha \bar{\nu} \mathcal{Z} \Delta \tilde{T} \nonumber
\end{gather}
where we have introduced
\begin{eqnarray}
    \mathbf{v}_R &=& \mathbf{V} - \epsilon\bar{r} \frac{\partial \hat{\mathbf{r}}}{\partial t} \\
    \mathbf{v}_E &=& \dot{\mathbf{X}} + \epsilon\bar{r}\frac{\partial \hat{\mathbf{r}}}{\partial t}
\end{eqnarray}
for notational convenience. The derivation of (\ref{kinetic_energy_equation}) can be found in Appendix B. 
\\\\ 
The symmetric first and second order kinetic energy equations are, respectively 
\begin{equation}
\label{1st_order_kinetic_energy_equation}
    \frac{w^{(0)}}{\sigma^{(0)}} \mathcal{H}^{(0)}_s + u_c^{(1)}\mathcal{H}^{(0)}_{\bar{r}}  = 0  
\end{equation}
and
\begin{gather}
u_c^{(2)}\mathcal{H}^{(0)}_{\bar{r}} + u_c^{(1)}\left(\mathcal{H}^{(1)}_c\right)_{\bar{r}} + \frac{w^{(0)}}{\sigma^{(0)}}\left(\mathcal{H}_c^{(1)}\right)_s  \nonumber \\
+ \frac{\sigma^{(1)}}{\sigma^{(0)}}u_c^{(1)}\mathcal{H}_{\bar{r}}^{(0)}  + \frac{w_c^{(1)}}{\sigma^{(0)}}\mathcal{H}_s^{(0)} = - \mathcal{H}^{(0)}_t + P^{(0)}_t   \nonumber \\ + \alpha T^{(0)}\Tilde{\mathcal{Z}}^{(0)}_t  - \left[\dot{\mathbf{X}}_s^{(0)}\cdot\hat{\tau}^{(0)}\right]\frac{\left(w^{(0)}\right)^2}{\sigma^{(0)}} \label{total_head_equation2}   \\+ \frac{\bar{\nu}}{\bar{r}}\left(\bar{r}w^{(0)}_{\bar{r}}\right)_{\bar{r}}w^{(0)}  + \frac{\bar{\nu}\mathcal{G}^{(0)}}{\bar{r}^2}\left(\mathcal{G}^{(0)}_{\bar{r}\bar{r}}-\frac{\mathcal{G}^{(0)}_{\bar{r}}}{\bar{r}}\right) \nonumber \\  + \frac{\alpha\bar{\eta}}{\bar{r}}\left(\bar{r}\tilde{T}_{\bar{r}}^{(0)}\right)_{\bar{r}} 
\Tilde{\mathcal{Z}}^{(0)}_c + \frac{\alpha}{2\bar{r}} \kappa^{(0)} z_b^{(0)} \mathcal{G}^{(0)} I \; \mathcal{H}^{(0)}_{\bar{r}} \nonumber
\end{gather} 
In section IV.C we will see that equation (\ref{total_head_equation2}) together with (\ref{symmetric_2nd_order_temperature_equation})  and (\ref{momentleadingorder_s}) yield the required evolution equations for $w^{(0)}$, $P^{(0)}$ and $T^{(0)}$.
\\\\
Through the introduction of the symmetrical stream function $\bar{M} = \mathcal{M}(\bar{r},s,t)$, we make the following coordinate transformation:
\begin{eqnarray}
    \left(\mathcal{M}, \text{id}\right): \mathbb{R}^3 \longrightarrow \mathbb{R}^3 \qquad\qquad\qquad \\
    (\bar{r},s,t) \mapsto (\bar{M},s,t) \qquad\qquad\qquad\\
    f_c(\bar{r},s,t) = \bar{f}_c(\bar{M},s,t) = f_c(\mathcal{M}(\bar{r},s,t),s,t)
\end{eqnarray}
where $f_c$ denotes any of the symmetrical field components; the variables $(\bar{M},s, t)$ are known as the von Mises variables. The inverse transformation is given by 
\begin{eqnarray}
    \left(\mathcal{R}, \text{id}\right) : \mathbb{R}^3 \longrightarrow \mathbb{R}^3 \\
    (\bar{M},s,t) \mapsto (\bar{r},s,t)
\end{eqnarray}
where ${\mathcal{R}} = \mathcal{M}^{-1}$. Note that ${\mathcal{R}}$ only exists at regular points $\mathbf{x}(\bar{r},s,t)$ where $\left(\frac{\partial\mathcal{M}}{\partial \bar{r}}\right)(\mathbf{x}) \neq 0$. \\\\
We define
\begin{eqnarray}
    M(\bar{r}, \varphi, s,t ; \epsilon) &=& 2\pi \int_0^{\bar{r}} w(\bar{r}, \varphi, s,t ; \epsilon)\bar{r}\;d\bar{r} \\
    &=& 2\pi \mathcal{M}(\bar{r}, \varphi, s,t ; \epsilon)
\end{eqnarray}
as the axial mass flux through a circular disk of radius $\bar{r}$. 
To see that $\mathcal{M}$ is indeed a stream function, observe that equation (\ref{symmetric_first_order_continutiy}) relates the leading order axial mass flux, $\mathcal{M}^{(0)}$, to the $\varphi$-averaged first order radial velocity, $u_c^{(1)}$, via the relations
\begin{subequations}
\label{leading_order_axial_mass_flux}
\begin{eqnarray}
\mathcal{M}^{(0)}(\bar{r},s,t) = \int_0^{\bar{r}} w^{(0)}r'\;dr' = \bar{M}\\
\sigma^{(0)}\bar{r} u_c^{(1)} = - \partial_s\mathcal{M}^{(0)}
\end{eqnarray}
\end{subequations}
Similarly, the symmetric second order continuity equation is satisfied by the introduction of the first order axial mass flux, $\mathcal{M}^{(1)}$:
\begin{subequations}
\label{first_order_axial_mass_flux}
\begin{gather}
    \sigma^{(0)} \bar{r}u_c^{(2)} + \sigma^{(1)}\bar{r}u_c^{(1)} + \frac{\bar{r}^2}{2}\left(\dot{\mathbf{X}}^{(0)}_s\cdot\hat{ \tau}^{(0)} \right) \qquad\qquad\\\nonumber \qquad\qquad-\frac{\sigma^{(0)}\kappa^{(0)}}{2}\alpha z_b^{(0)} \mathcal{G}^{(0)}I = - \left(\mathcal{M}_c^{(1)}\right)_s \\ \bar{r}w^{(1)}_c = \left(\mathcal{M}_c^{(1)}\right)_{\bar{r}}
\end{gather}
\end{subequations} \\\\
We can thus define the $s$-derivative along a stream line, $\bar{r} = R(\bar{M},s,t)$, at instant $t$ by $D_s = \partial_s +R_s\partial_{\bar{r}}$ where $R_s  = - \mathcal{M}^{(0)}_s/\mathcal{M}^{(0)}_{\bar{r}}= \sigma^{(0)}u_c^{(1)}/w^{(0)}$, and it is assumed that $w^{(0)}\neq 0$.  We additionally define the inverse operation  $\int_{\mathcal{S}}^{[0,s]}$, as an integration operator where $\bar{M}$ is held fixed. Consequently, the integration $g(\bar{M},s)= \int_{\mathcal{S}}^{[0,s]}f(\bar{M},s')\;ds'$ is the solution of $D_s(g)(\bar{M},s)=f(\bar{M},s)$. 
\subsection{Derivation of the evolution equations}
\label{section_IV_C}
Using the symmetric first  order circumferential momentum equation, and the first order kinetic energy equation, one can show that 
\begin{equation}
\label{compatability_conditions1}
    D_s\mathcal{G}^{(0)} = 0 \quad \text{ and } \quad D_s\mathcal{H}^{(0)} = 0
\end{equation}
Hence, given in terms of the von Mises variables, the leading order circulation and total head are $s$-independent, and we have 
\begin{eqnarray}
    \mathcal{G}^{(0)}(\bar{M},s,t) &=& \mathcal{G}^{(0)} (\bar{M}, t) \; = \;  \mathcal{G}^{(0)} (\bar{M}(\bar{r},s,t), t) \quad \\ 
    \mathcal{H}^{(0)}(\bar{M},s,t) &=& \mathcal{H}^{(0)}(\bar{M},t)\; = \;  \mathcal{H}^{(0)} (\bar{M}(\bar{r},s,t), t) \quad
\end{eqnarray}
In addition, from the symmetric first order temperature equation we immediately have
\begin{equation}
    D_sT^{(0)} = 0 
\end{equation}
and consequently 
\begin{equation}
    T^{(0)}(\bar{M},s,t) = T^{(0)}(\bar{M},t)= T^{(0)}(\bar{M}(\bar{r},s,t),t)
\end{equation}
We still need equations for the temporal evolution of $\mathcal{G}^{(0)}$, $\mathcal{H}^{(0)}$  and $T^{(0)}$, which are derived from the symmetric second order equations. \\\\
Using (\ref{leading_order_axial_mass_flux}),  (\ref{first_order_axial_mass_flux}) and (\ref{compatability_conditions1}) we can write the left hand side of (\ref{circulation_equation1}) as 
\begin{gather}
w^{(0)} D_s\left[\mathcal{G}_c^{(1)}-\mathcal{\mathcal{G}}^{(0)}_{\mathcal{M}^{(0)}}\mathcal{M}^{(1)}_c\right]-\frac{\bar{r}}{2}\mathcal{G}^{(0)}_{\bar{r}}\left(\dot{\mathbf{X}}^{(0)}_s\cdot\hat{ \tau}^{(0)}\right) \\ 
\nonumber \qquad\qquad + \frac{\sigma^{(0)}\kappa^{(0)}}{2\bar{r}}\alpha z_b^{(0)} \;I\; \mathcal{G}^{(0)}\mathcal{G}^{(0)}_{\bar{r}}
\end{gather}
Hence, we can eliminate the first order circulation, $\mathcal{G}_c^{(1)}$, by integrating  (\ref{circulation_equation1}) along a stream line. In the case of an open filament there will still be remaining boundary terms for $\mathcal{G}_c^{(1)}$, $\mathcal{M}_c^{(1)}$ and $\mathcal{G}_{\mathcal{M}^{(0)}}^{(0)}$. Concretely, this amounts to specifying the leading and first order velocity field components as well as the first order axial mass flux at the boundary. In the case of either a closed filament or an open filament which is symmetric along the stream line (as in  \citet{KleinTing1992}), we get the following evolution equation for the leading order circulation $\mathcal{G}^{(0)}$:

\begin{eqnarray}
\label{compatability_circulation}
  \int_{\mathcal{S}}^{[0,s]}\Bigg[-\mathcal{G}_t^{(0)}+ \frac{\bar{r}}{2\sigma^{(0)}}\mathcal{G}_{\bar{r}}^{(0)}\left(\dot{\mathbf{X}}^{(0)}_s \cdot\hat{\mathbf{\tau}}^{(0)}\right)\hspace{2cm}\\ \nonumber 
    + \bar{\nu} \left(\mathcal{G}^{(0)}_{\bar{r}\bar{r}}-\frac{\mathcal{G}_{\bar{r}}^{(0)}}{\bar{r}}\right)\Bigg]\frac{\sigma^{(0)}ds'}{w^{(0)}} = 0
\end{eqnarray}
where $\mathcal{S}$ denotes the integral over a streamline. \\
Note that the terms containing $I \; z_b^{(0)}$ (or equivalently \; $\psi_{12}^{(1)}$) which came from the Fourier component containing terms appearing in equation (\ref{second_ord_momentum_eq_theta}) have cancelled out.\\ 
\\
The temporal evolution equation for the leading order total head, $\mathcal{H}^{(0)}$, is given by (\ref{total_head_equation2}). 
The left hand side of (\ref{total_head_equation2}) equation has the same structure as the left hand side of (\ref{circulation_equation1}), hence we can employ the same trick and write the right hand side as:
\begin{gather}
\frac{w^{(0)}}{\sigma^{(0)}} D_s\left[\mathcal{H}_c^{(1)}-\mathcal{\mathcal{H}}^{(0)}_{\mathcal{M}^{(0)}}\mathcal{M}_c^{(1)}\right]-\frac{\bar{r}}{2\sigma^{(0)}}\mathcal{H}^{(0)}_{\bar{r}}\left(\dot{\mathbf{X}}^{(0)}_s\cdot\hat{ \tau}^{(0)}\right) \nonumber\\ 
 \qquad\qquad + \frac{\kappa^{(0)}}{2\bar{r}}\alpha z_b^{(0)} \;I\; \mathcal{G}^{(0)}\mathcal{H}^{(0)}_{\bar{r}}
\end{gather}
Thus, by integrating along a stream line, assuming an axially symmetric filament, we get an equation for the leading order total head, $\mathcal{H}^{(0)}\;$  \footnote{  Equation (\ref{compatability_total_head}) can also be derived in a different way without reference to the kinetic energy equation, but instead through multiplication of each component of the momentum equation at each order by associated components. Initially this was how we derived (\ref{compatability_total_head}), however, upon realisation that it may be simpler to use the kinetic equation, we then re-derived the equation. The former derivation thus becomes a cross check of this calculation, as the two derivations give the same result. This also explains why we have not introduced the kinetic equation in section III.A with the other equations: it is only when deriving equation (\ref{compatability_total_head}) that it makes sense to use and introduce the kinetic energy equation, which has, as far as we are aware, never been written before in the context of slender vortex expansions.}:
\begin{widetext}
\begin{eqnarray}
\label{compatability_total_head}
\int_{\mathcal{S}}^{[0,s]}\Bigg[
    - \mathcal{H}^{(0)}_t + P^{(0)}_t + \alpha \Tilde{T}^{(0)}\left(\Tilde{\mathcal{Z}}_c^{(0)}\right)_t + \left[\frac{\bar{r}\mathcal{H}^{(0)}_{\bar{r}}}{2\sigma^{(0)}}- \frac{\left(w^{(0)}\right)^2}{\sigma^{(0)}}\right]\dot{\sigma}^{(0)}  + \frac{\bar{\nu}}{\bar{r}}\left(\bar{r}w^{(0)}_{\bar{r}}\right)_{\bar{r}}w^{(0)} \\ + \frac{\bar{\nu}\mathcal{G}^{(0)}}{\bar{r}^2}\left(\mathcal{G}^{(0)}_{\bar{r}\bar{r}}-\frac{\mathcal{G}^{(0)}_{\bar{r}}}{\bar{r}}\right)+ \frac{\alpha\bar{\eta}}{\bar{r}}\left(\bar{r}\tilde{T}_{\bar{r}}^{(0)}\Tilde{\mathcal{Z}}^{(0)}_c\right)_{\bar{r}}\Bigg]\frac{\sigma^{(0)}ds'}{w^{(0)}} = 0 \nonumber
\end{eqnarray}
\end{widetext}
Again, we note that for an open filament which is not symmetric in some stream-wise direction, the boundary terms have to be specified. In particular, to fix the leading order total head at the boundary, in addition to specifying the leading and first order velocity components, the leading order pressure and temperature components in addition to the potential differential between the end points as determined by the integration over the streamline denoted by $\mathcal{S}$, must be specified. \\\\ The introduction of gravitational forces leads to a modification of the equations determining the leading order field components, in particular the leading order tangential velocity component, $w^{(0)}$. For the incompressible case analysed by  \citet{CallegariTing1978}, the authors were able to conclude that the effect of the tangential velocity is to reduce the velocity of the filament in the binormal direction. Thus, we would expect, depending on the sign of $\tilde{T}^{(0)}$ and $z_{\text{t}}^{(0)}$, that the presence of gravity either enhances or reduces this effect. 
\\\\
To close the system we need a third compatibility condition, namely a condition for the leading order temperature, $T^{(0)}$. Using the first and second order symmetric temperature equations, we get
can rewrite (\ref{symmetric_2nd_order_temperature_equation}) as 
\begin{gather}   \nonumber
    \sigma^{(0)} u_c^{(1)} \left(T_c^{(1)}\right)_{\bar{r}} + w^{(0)}\left(T_c^{(1)}\right)_s + w^{(1)}T_s^{(0)} + u_c^{(2)}T_{\bar{r}}^{(0)}  \\ + \sigma^{(1)}u_c^{(1)}T_{\bar{r}}^{(0)}   = - \sigma^{(0)}T_t^{(0)} + \frac{\sigma^{(0)}\bar{\eta}}{\bar{r}}\left(\bar{r}T_{\bar{r}}^{(0)}\right)_{\bar{r}} \quad  \\
    \nonumber + \frac{\sigma^{(0)}\kappa^{(0)}}{2\bar{r}} \alpha z_b^{(0)} \mathcal{G}^{(0)} I \; T^{(0)}_{\bar{r}} 
\end{gather}
Thus, it becomes clear that the left hand side can be written as
\begin{gather}
    w^{(0)}D_s\left[T_c^{(1)}-T^{(0)}_{\mathcal{M}^{(0)}}\mathcal{M}_c^{(1)}\right] - \frac{\bar{r}}{2}T_{\bar{r}}^{(0)}\dot{\mathbf{X}}^{(0)}_s\cdot\hat{\tau} \\ \nonumber \qquad \qquad
     + \frac{\sigma^{(0)}\kappa^{(0)}}{2\bar{r}}\alpha z_b^{(0)} \mathcal{G}^{(0)} \; I \; T_{\bar{r}}^{(0)}
\end{gather}
eventually yielding
\begin{eqnarray}
\label{compatability_temperature}
    \int_{\mathcal{S}}^{[0,s]}\Bigg[-T_t^{(0)} + \frac{\bar{\eta}}{\bar{r}}\left(\bar{r}T_{\bar{r}}^{(0)}\right)_{\bar{r}} \hspace{3cm}\\ \nonumber +\frac{\bar{r}}{2\sigma^{(0)}}T^{(0)}_{\bar{r}}\dot{\mathbf{X}}^{(0)}_s\cdot\hat{ \tau}^{(0)}\Bigg]\frac{\sigma^{(0)}ds'}{w^{(0)}} = 0
\end{eqnarray}
upon integration over a streamline $\mathcal{S}$, and assuming that the filament satisfies the symmetry condition.  \\
Thus, we have a complete set of equations for the motion of a buoyant vortex filament, namely (\ref{Coreconstant1}) and (\ref{coreconstant2}) together with the compatibility equations (\ref{compatability_circulation}), (\ref{compatability_temperature}), (\ref{compatability_total_head}).
\\\\
Currently, the variables in (\ref{compatability_temperature}), (\ref{compatability_total_head}) and (\ref{compatability_circulation}) are still given in terms of $(t,\bar{r},s)$. We wish to write them in terms on the von Mises variables $(t,\bar{M},s)$, in which $\mathcal{G}^{(0)}$, $\mathcal{H}^{(0)}$ and $T^{(0)}$ are $s$ independent. We can replace $\mathcal{G}^{(0)}_t$ by $\bar{\mathcal{G}}_t+\mathcal{G}_{\bar{M}}\bar{M}_t$ 
and the integral of the first term in  (\ref{compatability_circulation})  by 
\begin{equation}
\bar{\mathcal{G}}_t\int_{\mathcal{S}}^{[0,s]}\frac{\sigma^{(0)}}{w^{(0)}}ds' + \bar{\mathcal{G}}_{\bar{M}}\int_{\mathcal{S}}^{[0,s]}\mathcal{M}_t^{(0)}\frac{\sigma^{(0)}}{w^{(0)}}ds'
\end{equation}
to arrive at a differential equation for $\bar{\mathcal{G}}$. Here we have defined $\bar{f}$ as $f(\bar{r},s,t)=\bar{f}({\cal{M}}^{(0)}(\bar{r},s,t),s,t)$ to distinguish $f_t$ from $D_t(f)=\bar{f}_t$. We write $\bar{\mathcal{G}}_{\bar{M}}$ as above, even if there is no ambiguity for this derivative. We should note that in the case of horizontally placed filaments, the gravity terms drop out, and (\ref{compatability_circulation}) and (\ref{compatability_total_head}) reduce to those of \citet{KleinTing1992}. Furthermore, in this case one would be permitted to assume axial independence of the leading order field components, which would result in  (\ref{compatability_circulation}) and (\ref{compatability_total_head}) reducing to those of  \citet{CallegariTing1978}. 
\\\\
\section{Conclusions}
\label{sec:conclusions}
In summary, the equation of motion, valid to first order, is 
\begin{eqnarray}
\label{discussion_eq_of_motion}
 \dot{\mathbf{X}}^{(0)} =  \mathbf{Q}_0^{\perp} &+&\frac{\kappa^{(0)}}{4\pi}\left[\ln \frac{1}{\epsilon} + C_v(s,t) + C_w(s,t)\right]\hat{\mathbf{b}}^{(0)}\nonumber \\  
 &-& \frac{{1}}{4\pi}  \alpha C_T(s,t) \hat{\mathbf{z}}\times \hat{\mathbf{\tau}}^{(0)}
\end{eqnarray}
where $C_v(s,t)$, $C_w(s,t)$ and $C_T(s,t)$ are given by  (\ref{Core_constant}), (\ref{Core_constant_TingKlein}) and (\ref{Core_constant_gravity_part}), while {$\alpha = -\beta T_0/\bar{\lambda}^2$ with $\beta$ and $T_0$ defined in (\ref{defintion_beta_T_0})}. {Coming back to dimensional variables, using provided definitions of $\epsilon$, $\alpha$, Re and Fr, yields
\begin{eqnarray}
\label{buoyancy_eq_of_motion_dim}
 \dot{\mathbf{X}}^{(0)} =  \mathbf{Q}_0^{\perp} &+&\frac{\Gamma\kappa^{(0)}}{4\pi}\left[\ln \frac{1}{\epsilon} + \tilde{C}_v(s,t) + \tilde{C}_w(s,t)\right]\hat{\mathbf{b}}^{(0)}\nonumber \\  
 &+& \frac{1}{4\pi} \frac{g}{\Gamma}\beta  C_T(s,t) \hat{\mathbf{z}}\times \hat{\mathbf{\tau}}^{(0)} 
\end{eqnarray}
where 
\begin{eqnarray}
\label{Core_constant_gravity_part}
   C_T(s,t) = -8\pi^2
    \int_0^{\infty}\xi\tilde{T}^{(0)}\; d\xi\nonumber
\end{eqnarray}
and where we have written 
\begin{eqnarray}
	\label{Core_constant_dim}
	\tilde{C}_v(s,t) &=& \lim_{\bar{r}\longrightarrow \infty}\left(\frac{4\pi^2}{\Gamma^2}\int_0^{\bar{r}}\xi\left(v^{(0)}\right)^2\;d\xi -\ln\bar{r}\right) +\frac{1}{2} \nonumber\\
\\
	\label{Core_constant_TingKlein_dim}
	\tilde{C}_w(s,t) &=& -\frac{8\pi^2}{\Gamma^2}\int_0^{\infty}\xi\left(w^{(0)}\right)^2\;d\xi
\end{eqnarray}
to be in accordance with the notation of \citet{CallegariTing1978}, and to indicate that all of the field components are now dimensional quantities. To check the dimensions of the buoyancy term, we may note that since $\beta$ is the inverse of a temperature, this term has the dimension of $g l^2 /\Gamma$ which is a velocity.}
The {filament} equation of motion{, (\ref{discussion_eq_of_motion}), or equivalently (\ref{buoyancy_eq_of_motion_dim}),} couples to the compatibility conditions,  (\ref{compatability_circulation}), (\ref{compatability_total_head}) and (\ref{compatability_temperature}), to form a complete set of equations for the motion of a buoyant vortex filament assuming non-vanishing leading order axial velocity. Equation (\ref{discussion_eq_of_motion}) is equivalent to the equation obtained by  \citet{ChangSmith2018} for a filament with continuous density variation within the core, excluding higher order terms. {A comparison to the results of \citet{Saffmann1993} for buoyant circular vortex rings is given in Appendix D.} The perhaps most interesting feature of (\ref{discussion_eq_of_motion}) is that it now contains a contribution in the normal direction arising from the motion inside the core. It is this term that results in the expansion of buoyant vortex rings as they rise, as was demonstrated by \citet{ChangSmith2018}. The binormal part of the additional bouyancy term appearing in (\ref{discussion_eq_of_motion}), is responsible for the tendency of bouyant vortex rings to align themselves in the horizontal plane. 
Once aligned in the horizontal plane gravity, by the compatibility conditions, ceases to have an effect on the leading order total head, and only enters directly in the core constant. Furthermore, the introduction of gravity at first order necessarily imposes the condition of axial dependence on the leading order field components. For a nearly straight, vertically aligned vortex filament the influence of gravity will largely be indirectly through the compatability conditions, in particular through its influence on the leading order tangential velocity component. However, once it is slightly tilted it will also necessarily have another contribution coming from the core constant. This contribution would become more significant the greater the tilt. Both of these contributions would in addition have to compete with the logarithmic term in  (\ref{discussion_eq_of_motion}), which for truly thin filaments can be much larger. Note, however, that for say, $\epsilon = 0.1$, for which the  asymptotic predictions are relatively accurate according to \citet{KleinKnio1995}, we have $\ln(1/\epsilon)\sim 2$, so that the logarithmic term does not clearly dominate the terms of order unity. \\\\
The next step would be to allow the background density to vary with height $z$, following the anelastic approximation. One could also go one step further and employ the pseudo-incompressible approximation for low Mach number flows. We are certain such an analysis would yield new and interesting results, bringing the theory closer to an accurate description of the dynamics of tornado-like vorticies. \\

\begin{acknowledgments}
This research was conducted while Marie Rodal was employed at FB Mathematik and Informatik at Freie Univeristät Berlin, and was funded by the Deutsche 
Forschungsgemeinschaft (DFG) through grant CRC 1114 "Scaling Cascades in Complex Systems", Project Number 235221301, Project B07: Self-similar structures in turbluent flows and the construction of LES closures. 

Marie Rodal additionally acknowledges with gratitude the
support of the Research Fund of the University of Antwerp.
\\\\
Finally, we  thank the developers of SageMath\citep{sagemath}, an open source computer algebra system (or symbolic calculator) with which we performed symbolic calculation to cross check and chase typos and errors in such a complex calculation.
\end{acknowledgments}

\section*{Code Availability}
The SageMath code containing the calculations of the preceding sections can be found in the dedicated GitHub repository at https://github.com/danielmargerit/SageMAE4PDEs.

\section*{Declaration of interests}
The authors report no conflict of interest.

\section{Appendix}
\appendix
\section{Computation of core constant integral}
We wish to simplify the integral 
\begin{eqnarray}
C_0(s,t) = \lim_{\bar{r}\longrightarrow\infty} \Bigg[
\frac{v^{(0)}\kappa^{(0)} }{\bar{r}}\int_0^{\bar{r}} \frac{1}{2\left(v^{(0)}\right)^2}\int_0^z \left[\xi v^{(0)} H(\xi,t)d\xi\right]dz \nonumber  \\
- \frac{{1}}{4\pi}\ln \bar{r}\Bigg] \qquad
\end{eqnarray}
where 
\begin{equation}
    H(\xi,t) = 2\xi \zeta^{(0)} + v^{(0)} + \frac{2 \xi w^{(0)}w^{(0)}_{\bar{r}}}{v^{(0)}}
\end{equation}
and the asymptotic behaviour of $v^{(0)}$, $\zeta^{(0)}$ and $w^{(0)}$ are given by (\ref{boundcond_v0}), (\ref{boundcond_w0}) and (\ref{leading_order_circulation}). We note that this calculation was first carried out by \citet{CallegariTing1978}, but as these authors did not explicitly include all the details of the calculation, we are including them here to avoid any potential confusion.\\\\
To simplify, we write 
\begin{equation}
\label{appendix_core_constant}
    C_0(s,t) = \kappa^{(0)} \lim_{\bar{r}\longrightarrow \infty} \left[ \frac{v^{(0)}}{\bar{r}} h(\bar{r},t) - \frac{{1}}{4\pi} \ln\bar{r}\right]
\end{equation}
where we have defined
\begin{equation}
    h(\bar{r},t) = \int_0^{\bar{r}} \frac{1}{z \left(v^{(0)}\right)^2}\int_0^z \left[\xi v^{(0)}H(\xi,t) d\xi\right]dz    
\end{equation}
Consequently, we get 
\begin{eqnarray}
    C_0(s,t) &=& \kappa^{(0)} \lim_{\bar{r}\longrightarrow \infty} \left[\frac{v^{(0)}\bar{r}}{\bar{r}^2}h(\bar{r},t) - \frac{{1}}{4\pi}\ln \bar{r}\right] \\
    &=& \kappa^{(0)} \lim_{\bar{r}\longrightarrow\infty}\left[\frac{{1}}{2\pi \bar{r}^2}h(\bar{r},t) - \frac{{1}}{4\pi} \ln \bar{r}\right] \\
    &=& \frac{{1}}{2 \pi} \kappa^{(0)}\lim_{\bar{r}\longrightarrow\infty} \left[\frac{h(\bar{r},t)-\frac{\bar{r}^2}{2}\ln \bar{r}}{\bar{r}^2}\right]
\end{eqnarray}
Application of the l'Hopital's rule, yields
\begin{eqnarray}
\nonumber
    C_0(s,t) = \frac{{1}}{2\pi} \kappa^{(0)} \lim_{\bar{r}\longrightarrow\infty}\left[\frac{\left(h-\frac{\bar{r}^2}{2}\ln \bar{r}\right)'}{\left(\bar{r}^2\right)'}\right] \qquad\qquad\qquad\qquad\\
    \nonumber
    = \frac{{1}}{2\pi}\kappa^{(0)}\lim_{\bar{r}\longrightarrow\infty} \left[ \frac{\frac{1}{\bar{r}\left(v^{(0)}\right)^2}\int_0^{\bar{r}} \left[\xi v^{(0)}H(\xi,t)d\xi\right]-\frac{\bar{r}}{2}-\bar{r}\ln\bar{r}}{2\bar{r}}\right] \\\nonumber
    = \frac{{1}}{4\pi}\kappa^{(0)} \lim_{\bar{r}\longrightarrow\infty}\left[\frac{4\pi^2}{{1}^2}\int_0^{\bar{r}} \xi v^{(0)} H(\xi,t) d\xi - \frac{1}{2} - \ln{\bar{r}}\right] \qquad\;\;\\\nonumber
    = \frac{{1}}{4\pi}\kappa^{(0)} \lim_{\bar{r}\longrightarrow\infty} \left[\int_0^{\bar{r}}\xi v^{(0)} H(\xi,t) d\xi - \frac{1}{2} -\ln\bar{r}\right]\nonumber\qquad\qquad\;\;
\end{eqnarray}
Next, we note that 
\begin{eqnarray}
   \lim_{\bar{r}\longrightarrow\infty} \int_0^{\bar{r}} \xi v^{(0)} \left(\frac{2\xi w^{(0)}w^{(0)}_{\bar{r}}}{v^{(0)}}\right) d\xi  \qquad\qquad\qquad\qquad\\\
    =  \lim_{\bar{r}\longrightarrow\infty}2\int_0^{\bar{r}}\xi^2 w^{(0)}w^{(0)}_{\bar{r}} d\xi \qquad\qquad\qquad\quad\\
    =  \lim_{\bar{r}\longrightarrow\infty}\int_0^{\bar{r}} \xi^2 \frac{\partial}{\partial \xi}\left(w^{(0)}\right)^2 d\xi \qquad\qquad\qquad\\
    = \left[\xi^2 \left(w^{(0)}\right)^2\right]_0^{\infty}-\int_0^{\infty} \left(w^{(0)}\right)^2 2\xi d\xi  \quad\quad\\
    = -2 \int_0^{\infty}\left(w^{(0)}\right)^2 \xi d\xi \;\qquad\qquad\qquad\qquad
\end{eqnarray}
Similarly, we have
\begin{eqnarray}
    \int_0^{\bar{r}} 2\xi^2 v^{(0)}\zeta^{(0)} d\xi &=& \int_{0}^{\bar{r}} 2\xi v^{(0)}\left(\xi v^{(0)}\right)_\xi d\xi \\
    &=&\int_0^{\bar{r}} \frac{\partial}{\partial\xi}\left(\xi v^{(0)}\right)^2 d\xi \\
    &=& \left[\left(\xi v^{(0)}\right)^2\right]^{\bar{r}}_0 \\
    &=& \left(\bar{r}v^{(0)}\right)^2
\end{eqnarray}
and thus, 
\begin{eqnarray}
\nonumber
    \lim_{\bar{r}\longrightarrow\infty} 4\pi^2\left(\int_0^{\bar{r}}2\xi^2 v^{(0)}\zeta^{(0)}d\xi\right) 
    &=& \lim_{\bar{r}\longrightarrow\infty} 4\pi^2 \left(\bar{r}v^{(0)}\right)^2  \\\nonumber
    &=& 4\pi^2 \left(\frac{{1}}{2\pi}\right)^2 \\\nonumber
    &=& 1
\end{eqnarray}
Consequently, equation $(\ref{appendix_core_constant})$ can be written as
\begin{eqnarray}
\label{CT_core_constant}
    C_0(s,t) = \frac{{1}}{4\pi} \kappa^{(0)}\Bigg[\lim_{\bar{r}\longrightarrow\infty} \left(4\pi^2 \int_0^{\bar{r}} \xi \left(v^{(0)}\right)^2d\xi-\ln\bar{r}\right)\qquad \\ +\frac{1}{2} - 8\pi^2\int_0^\infty \left(w^{(0)}\right)^2\xi d\xi\Bigg] \qquad \nonumber
\end{eqnarray}
\section{The kinetic energy equation}
The Navier Stokes equation in Cartesian coordinates given the Boussinesq approximation is 
\begin{equation}
    \left(\frac{\partial\mathbf{v}}{\partial t}\right)_{xyz}
    +\mathbf{v}\cdot\nabla_{xyz}\mathbf{v} = - \nabla_{xyz} P + \bar{\nu}\epsilon^2 \Delta_{xyz}\mathbf{v} - \epsilon^{-2}\alpha\tilde{T}\hat{\mathbf{z}}
\end{equation}
where we use the subscript $xyz$ to indicate that these are derivatives with respect to the Cartesian coordinate system. \\
The kinetic energy equation is obtained by contracting both sides with $\mathbf{v}$, yielding
\begin{eqnarray}
\label{kinetic_energy_eqn_velocity}
\left(\frac{\partial (\mathbf{v}^2/2)}{\partial t}\right)_{xyz} = - \mathbf{v} \cdot\nabla_{xyz} \left[\mathbf{v}^2/2 +  P\right]  + \bar{\nu}\epsilon^2\mathbf{v}\cdot\Delta_{xyz}\mathbf{v}\qquad \\  - \epsilon^{-2}\alpha\tilde{T}\mathbf{v}\cdot\hat{\mathbf{z}}  \nonumber  
\end{eqnarray}
where we have used 
\begin{eqnarray}
    \nabla_{xyz}(\mathbf{v}^2/2) &=& \mathbf{v}\cdot\nabla_{xyz}\mathbf{v} + \mathbf{v}\times\mathbf{w}
\end{eqnarray}
and
\begin{equation}
    \mathbf{v}\cdot\left(\mathbf{v}\times \mathbf{w}\right) = 0
\end{equation}
We can write (\ref{kinetic_energy_eqn_velocity}) in terms of the total head, $\mathcal{H}$, in Cartesian coordinates, as follows 
\begin{gather}
\label{kinetic_energy_equation_total_head}
    \left(\frac{\partial\mathcal{H}}{\partial t}\right)_{xyz} + \mathbf{v}\cdot\nabla_{xyz} \mathcal{H} = \left(\frac{\partial P}{\partial t}\right)_{xyz} + \epsilon^2 \bar{\nu}\mathbf{v}\cdot \Delta_{xyz} \mathbf{v}  \\ + \epsilon^{-2}\alpha T\left(\frac{\partial \mathcal{Z}}{\partial t}\right)_{xyz} \nonumber
    + \alpha \bar{\eta}\mathcal{Z}\Delta_{xyz}T
\end{gather}
where 
\begin{equation}
    \mathcal{H} = P + \mathbf{v}^2/2 + \epsilon^{-2} \alpha \tilde{T}\mathcal{Z}
\end{equation}
$\mathcal{Z} = z - z_0$ is the height and 
\begin{equation}
    \nabla_{xyz}\mathcal{Z} = \hat{\mathbf{z}} 
\end{equation}
In deriving (\ref{kinetic_energy_equation_total_head}) we have used the temperature equation, which in Cartesian coordinates is
\begin{equation}
    \left(\frac{\partial T}{\partial t}\right)_{xyz} + \mathbf{v}\cdot\nabla_{xyz} T = \epsilon^2 \bar{\eta} \Delta_{xyz} T
\end{equation}
to remove the time derivatives of the temperature $T$. 
In curvilinear coordinates (\ref{kinetic_energy_equation_total_head}) becomes 
\begin{gather}
\left(\frac{\partial \mathcal{H}}{\partial t}\right)_{r\varphi s} + \mathbf{v}_R\cdot\nabla_{r\varphi s}\mathcal{H} = \left(\frac{\partial P}{\partial t}\right)_{r \varphi s} - \mathbf{v}_E \cdot\nabla_{r\varphi s} P \nonumber \\\label{kinetic_energy_equation_total_head_curvilinear_coord}   + \epsilon^2\bar{\nu}\left[\dot{\mathbf{X}}(s,t) + \mathbf{V}\right]\cdot\Delta_{r\varphi s}\left[\dot{\mathbf{X}}(s,t) + \mathbf{V}\right] + \alpha \bar{\eta} \mathcal{Z} \Delta_{r\varphi s} T \\ \nonumber +\epsilon^{-2}\alpha T \left(\frac{\partial \mathcal{Z}}{\partial t}\right)_{r \varphi s}   -\epsilon^{-2}\alpha T \mathbf{v}_E\cdot\nabla_{r\varphi s}\mathcal{Z} 
\end{gather}
where 
\begin{eqnarray} 
    \mathbf{v}_R &=& \mathbf{V}- r\left(\frac{\partial\hat{\mathbf{r}}}{\partial t}\right)_{r\varphi s} \\
    \mathbf{v}_E &=& \dot{\mathbf{X}} + r\left(\frac{\partial \hat{\mathbf{r}}}{\partial t}\right)_{r\varphi s} 
\end{eqnarray}
and
\begin{gather}
    \mathcal{H} = P + \left(\dot{\mathbf{X}}(s,t) + \mathbf{V}\right)^2/2 +\epsilon^{-2}\alpha\tilde{T} \mathcal{Z} \\
    \mathcal{Z}(\bar{r}, \varphi, s,t) =
    \left[\mathbf{X}(s,t) + r\hat{\mathbf{r}} - \mathbf{X}(0,t)\right]\cdot\hat{\mathbf{z}} 
\end{gather}
To transform equation (\ref{kinetic_energy_eqn_velocity}) we have used 
\begin{gather}
    \left(\frac{\partial \mathcal{H}}{\partial t}\right)_{xyz} + \mathbf{v}\cdot \nabla_{xyz} \mathcal{H} = \left(\frac{\partial \mathcal{H}}{\partial t}\right)_{r\varphi s} + \mathbf{v}_R \cdot\nabla_{r\varphi s} \mathcal{H} \qquad\qquad \\
\left(\frac{\partial P}{\partial t}\right)_{xyz} = \left(\frac{\partial P}{\partial t}\right)_{r \varphi s} - \mathbf{v}_E \cdot \nabla_{r\varphi s} P \qquad\\    
\left(\frac{\partial \mathcal{Z}}{\partial t}\right)_{xyz} = \left(\frac{\partial \mathcal{Z}}{\partial t}\right)_{r \varphi s} - \mathbf{v}_E \cdot \nabla_{r \varphi s} \mathcal{Z}\qquad
\end{gather}
Here $\nabla_{r \varphi s} \mathcal{Z} = \mathbf{\hat{z}}$. 
More details on the coordinate transformation can be found in \citet{CallegariTing1978}, Appendix A and B. 
\\\\
The leading and second order equations of (\ref{kinetic_energy_equation_total_head_curvilinear_coord}) are, respectively 
\begin{equation}
    u^{(1)}\mathcal{H}^{(0)}_{\bar{r}} + \frac{v^{(0)}}{\bar{r}}\mathcal{H}_\varphi^{(1)} + \frac{\omega^{(0)}}{\sigma^{(0)}} \mathcal{H}_s^{(0)}
    =-\dot{\mathbf{X}}^{(0)}\cdot\hat{\mathbf{r}}^{(0)}P^{(0)}_{\bar{r}}
\end{equation}
and
\begin{gather}
    \mathcal{H}_t^{(0)} + u^{(2)}\mathcal{H}^{(0)}_{\bar{r}} + u^{(1)}\mathcal{H}^{(1)}_{\bar{r}} + \frac{v^{(1)}}{\bar{r}}\mathcal{H}^{(1)}_\varphi \nonumber\\ + \frac{w^{(1)}}{\sigma^{(0)}}\mathcal{H}^{(0)}_s  + \frac{v^{(0)}}{\bar{r}} \mathcal{H}^{(2)}_\varphi + \frac{w^{(0)}}{\sigma^{(0)}}\mathcal{H}^{(1)}_s - \frac{w^{(0)}\sigma^{(1)}}{\left(\sigma^{(0)}\right)^2}\mathcal{H}^{(0)}_s \nonumber\\  = P_t^{(0)} - \left(\dot{\mathbf{X}}\cdot \hat{\mathbf{r}}\right)^{(1)}P_{\bar{r}}^{(0)} - \dot{\mathbf{X}}^{(0)}\cdot \hat{\mathbf{r}}^{(0)}P^{(1)}_{\bar{r}} \nonumber\\ - \frac{1}{\bar{r}} \dot{\mathbf{X}}^{(0)}\cdot\hat{\theta}^{(0)} P^{(1)}_\varphi   + \alpha \Tilde{T}^{(0)} \mathcal{Z}^{(0)}_t  + \bar{\nu} \mathbf{V}^{(0)}\cdot\Delta \mathbf{V}^{(0)} \\ \nonumber - \alpha \Tilde{T}^{(0)}\mathbf{v}_E^{(0)}\cdot \mathbf{\hat{z}} + \alpha\bar{\eta}\mathcal{Z}^{(0)}\Delta \tilde{T}^{(0)} \nonumber
\end{gather}
where we have dropped the subscript ($\bar{r}\varphi s$) for notational convenience. 
Averaging with respect to $\varphi$ yields the leading and first order symmetric equations:
\begin{equation}
    \frac{w^{(0)}}{\sigma^{(0)}}\mathcal{H}^{(0)}_s + u_c^{(1)}\mathcal{H}_{\bar{r}}^{(0)} = 0 
\end{equation}
and 
\begin{gather}\nonumber
    \mathcal{H}^{(0)}_t + u_c^{(2)}\mathcal{H}^{(0)}_{\bar{r}} + u_c^{(1)}\left(\mathcal{H}^{(1)}_c\right)_{\bar{r}} 
    + \frac{w^{(1)}_c}{\sigma^{(0)}}\mathcal{H}_s^{(0)} + \\ \frac{w^{(0)}}{\sigma^{(0)}} \nonumber\left(\mathcal{H}_c^{(1)}\right)_s - \frac{w^{(0)}\sigma^{(1)}}{\left(\sigma^{(0)}\right)^2} \mathcal{H}^{(0)}_s + 
    \frac{1}{2}\left(\mathcal{H}_{11}^{(0)}\right)u^{(1)}_{11} \\+ \frac{1}{2}\left(\mathcal{H}_{12}^{(1)}\right)_{\bar{r}}u^{(1)}_{12} - \frac{1}{2\bar{r}}\mathcal{H}_{11}^{(1)} v_{12}^{(1)} + \frac{1}{2\bar{r}}\mathcal{H}_{12}^{(1)}v_{11}^{(1)} \\ \nonumber
    = P_t^{(0)} - \frac{1}{2}\dot{\mathbf{X}}^{(0)} \cdot \hat{\mathbf{n}}^{(0)} \left(P^{(1)}_{11}\right)_{\bar{r}} - 
    \frac{1}{2} \dot{\mathbf{X}}^{(0)} \cdot \hat{\mathbf{b}}^{(0)} \left(P_{12}^{(1)} \right)_{\bar{r}}\\ \nonumber  - \frac{1}{2\bar{r}}\dot{\mathbf{X}}^{(0)} \cdot \hat{\mathbf{b}}^{(0)} P_{12}^{(1)}
    -\frac{1}{2\bar{r}} \dot{\mathbf{X}}^{(0)} \cdot \hat{\mathbf{n}}^{(0)}P_{11}^{(1)} -\alpha \Tilde{T}^{(0)} \dot{\mathbf{X}}^{(0)}\cdot \hat{\mathbf{z}}  \\ + \alpha\Tilde{T}^{(0)}\left(\mathcal{Z}_c^{(0)}\right)_t + \bar{\nu} \mathbf{V}^{(0)} \cdot\Delta \mathbf{V}^{(0)} + \alpha \bar{\eta} \mathcal{Z}_c^{(0)} \Delta \Tilde{T}^{(0)} \nonumber
\end{gather}
Here we have written the asymmetric part of $\mathcal{H}^{(1)}$ as
\begin{gather}
    \mathcal{H}_a^{(1)} = \mathcal{H}^{(1)}_{11}\cos \varphi + \mathcal{H}^{(1)}_{12}\sin \varphi
\end{gather}
where
\begin{gather}
    \mathcal{H}_{11}^{(1)} = P_{11}^{(1)} + v^{(0)} v_{11}^{(1)} + w^{(0)}w_{11}^{(1)}\qquad\qquad \\ \qquad\qquad+ \alpha \mathcal{Z}_c^{(0)}T_{11}^{(1)} + \alpha \bar{r} \tilde{T}^{(0)}z_{\text{n}}^{(0)} + v^{(0)} \dot{\mathbf{X}}^{(0)} \cdot \hat{\mathbf{b}}^{(0)} \nonumber \\
    \mathcal{H}_{12}^{(1)} = P_{12}^{(1)} + v^{(0)} v_{12}^{(1)} + w^{(0)}w_{12}^{(1)}\qquad\qquad \\ \qquad\qquad + \alpha \mathcal{Z}_c^{(0)}T_{12}^{(1)}+ \alpha \bar{r} \tilde{T}^{(0)}z_{\text{b}}^{(0)} + v^{(0)} \dot{\mathbf{X}}^{(0)} \cdot \hat{\mathbf{b}}^{(0)}\nonumber
\end{gather}
We additionally note that, using (\ref{assymetric_1st_order_momentum_eqn_v}) we can write $P_{11}^{(1)}$ and $P_{12}^{(1)}$ as
\begin{gather}
    P_{11}^{(1)} = u_{12}^{(1)} \bar{r}v_{\bar{r}}^{(0)} -v^{(0)}v_{11}^{(1)} + v^{(0)}u^{(1)}_{12}\qquad\qquad\qquad 
\\ \qquad\qquad\qquad - \left(w^{(0)}\right)^2 \nonumber \kappa^{(0)} - \alpha \bar{r} \Tilde{T}^{(0)}z_{\text{n}}^{(0)} \\
P_{12}^{(1)} = -u_{11}^{(1)} \bar{r}v_{\bar{r}}^{(0)} - v^{(0)} u_{11}^{(1)} -v^{(0)}v_{12}^{(1)} - \alpha \bar{r} \Tilde{T}^{(0)} z_{\text{b}}^{(0)}
\end{gather}
Using these relations, (B23) can be rewritten to eventually yield (\ref{total_head_equation2}). 
\section{The case with no leading order axial velocity}
When deriving the evolution equations in section IV, the assumption was made that $w^{(0)}\neq 0$ at any point in the vortex core. This allowed us to introduce the s-derivative along stream lines, and thus conclude that along stream lines the leading order circulation, total head and temperature are all s-independent on these curves. However, the case when $w^{(0)}=0$ at any point in the vortex core requires special attention, which we will briefly cover in this section. We will not treat here the special case of a stagnation point where $w^{(0)}=0$ at a single point. 
\\\\ Whenever $w^{(0)}=0$ we get, from (\ref{assymetric_1st_order_momentum_eqn_w}), (\ref{symmetric_first_order_continutiy}) and (\ref{bc_at_r=0})
\begin{equation}
    w_a^{(1)} = 0 \qquad \text{ and } \qquad u_c^{(1)} = 0  
\end{equation}
Using this in (\ref{symmetric_first_order_moment_eq_s}) and (\ref{second_ord_momentum_eq_s}) yields
\begin{eqnarray}
    P_s^{(0)} &=& - \alpha\sigma^{(0)} \tilde{T}^{(0)} z_{\text{t}}^{(0)} \\
    P_s^{(1)} &=& -\alpha\sigma^{(0)}\left(\tilde{T}^{(1)}_c z_{\text{t}}^{(0)}+ \tilde{T}^{(0)}z_t^{(1)}
    + \frac{\sigma^{(1)}}{\sigma^{(0)}}\Tilde{T}^{(0)}z_{\text{t}}^{(0)}\right)  \qquad
\end{eqnarray}
Thus, if $P_s^{(0)} = 0$, then we must necessarily have that $\Tilde{T}^{(0)} = 0$ and we revert back to the case described by \citet{CallegariTing1978} with no gravitational force and no leading order axial variation. \\\\ Assuming $w^{(0)}=0$ and $P_s^{(0)}\neq 0$, the symmetric second order circumferential momentum and temperature equations are, respectively  
\begin{gather}
v_t^{(0)} + u_c^{(2)}\frac{1}{\bar{r}}(\bar{r}v^{(0)})_{\bar{r}} + \frac{w_c^{(1)}v_s^{(0)}}{\sigma^{(0)}} \qquad\qquad\nonumber \\ = \frac{\bar{\nu}}{\bar{r}}\left(\bar{r}v^{(0)}_{\bar{r}}\right)_{\bar{r}} - \frac{\bar{\nu}}{\bar{r}}v^{(0)} + \frac{\alpha}{2}\kappa^{(0)}z_b^{(0)}v^{(0)}I \zeta^{(0)}
\label{appendix_C_circum_mom_second_order}
\end{gather}
and
\begin{gather}
    T_t^{(0)} + u_c^{(2)}T^{(0)}_{\bar{r}} + \frac{w_c^{(1)}}{\sigma^{(0)}} T_s^{(0)} \qquad\qquad\nonumber \\ = \frac{\bar{\eta}}{\bar{r}}\left(\bar{r}T_{\bar{r}}^{(0)}\right)_{\bar{r}} + \frac{\alpha }{2}\kappa^{(0)}z_b^{(0)} v^{(0)} I T_{\bar{r}}^{(0)}
    \label{appendix_C_temperature_second_order}
\end{gather}
From the second order continuity equation, we additionally have 
\begin{gather}
    u_c^{(2)} = -\frac{\bar{r}}{2\sigma^{(0)}} \left(\dot{\mathbf{X}}_s^{(0)}\cdot \hat{\tau}^{(0)}\right) + \frac{\kappa^{(0)}}{2}\alpha z_b^{(0)} v^{(0)} I \qquad\qquad\nonumber\\
    - \frac{1}{\bar{r}\sigma^{(0)}}\int_0^{\bar{r}} r' \left(w_c^{(1)}\right)_s dr'
\end{gather}
Contrary to the case without axial variation, here $w_c^{(1)}$ appears in those leading order equation for the core structure dynamical evolution, which means that an equation for $w_c^{(1)}$ is needed to close the equations.
The third order symmetric tangential momentum equation yields the temporal evolution equation for $w_c^{(1)}$:
\begin{gather}
    (w_c^{(1)})_t  + g_4^{(3)} w_c^{(1)} + u_c^{(2)}\left(w_c^{(1)}\right)_{\bar{r}} + \frac{w_c^{(1)}}{\sigma^{(0)}} \left(w_c^{(1)}\right)_s \nonumber \\ =  - \frac{\left(P_c^{(2)}\right)_s}{\sigma^{(0)}} +\frac{\sigma^{(1)}}{\left(\sigma^{(0)}\right)^2}\left(P_c^{(1)}\right)_s + \frac{\bar{\nu}}{\bar{r}}\left(\bar{r}\left(w_c^{(1)}\right)_{\bar{r}}\right)_{\bar{r}} \\ 
    - \alpha \left(\tilde{T}_c^{(2)} z_{\text{t}}^{(0)} + \tilde{T}_c^{(1)}z_\text{t}^{(1)} + \tilde{T}^{(0)}z_{\text{t}}^{(2)}\right) + s_4^{(0)} \nonumber
\end{gather}
where we have defined
\begin{eqnarray}
    g_4^{(3)} = \frac{\dot{\mathbf{X}}^{(0)}_s\cdot \hat{\tau}^{(0)}}{\sigma^{(0)}} - \frac{\kappa^{(0)}}{2}\left(u_{11}^{(1)}-v_{12}^{(1)}\right)
\end{eqnarray}
and
\begin{gather}
    s_4^{(3)} = -\ddot{\mathbf{X}}^{(0)}\cdot\hat{\tau}^{(0)}  - \kappa^{(0)}v^{(0)}\left<w_a^{(2)}\sin\varphi\right> - \frac{\kappa^{(0)}\bar{r}}{2\sigma^{(0)}}\left(P_{11}^{(1)}\right)_s \nonumber\\ 
    -\frac{1}{2}\frac{\partial\hat{\mathbf{n}}^{(0)}}{\partial t} \cdot\hat{\tau}^{(0)}\left( u_{11}^{(1)}-v_{12}^{(1)}\right)  -\frac{1}{\bar{r}}\left<v_a^{(1)}\left(w_a^{(2)}\right)_\varphi\right>
    \\ \nonumber
    -\frac{1}{2}\frac{\partial\hat{\mathbf{b}}^{(0)}}{\partial t} \cdot\hat{\tau}^{(0)}\left(u_{12}^{(1)} + v_{11}^{(1)} -v^{(0)}\kappa^{(0)}\bar{r} \right)-\left<u_a^{(1)}\left(w_a^{(2)}\right)_{\bar{r}}\right> \\ \nonumber
    + \frac{\sigma^{(2)}\sigma^{(0)} - \left(\sigma^{(1)}\right)^2-\frac{1}{2}\left(\kappa^{(0)}\sigma^{(0)}\bar{r}\right)^2}{\left(\sigma^{(0)}\right)^3} P_s^{(0)} - \frac{\bar{r}\kappa^{(0)}\mathcal{T}_{\text{o}}^{(0)}}{2}P_{12}^{(1)}
\end{gather}\\
Here we have introduced the notation $\left<...\right>$ to indicate averaging w.r.t. $\varphi$. The $P_c^{(2)}$-term in (C7) can be removed by averaging over $s$, while $w_a^{(2)}$ can be found from the asymmetric part of the 2nd order tangential momentum equation (not shown). However, $P_c^{(1)}$ is related to $v_c^{(1)}$ through (\ref{symmetric_first_order_moment_eq_r}), and thus a fourth evolution equation, for $v_c^{(1)}$, is required. This equation is given by the symmetric 3rd order circumferential momentum equation (not shown). The complete derivation lies outside the scope of this work, as it will require the higher order analysis (i.e. derivation of expressions for $u_a^{(2)}$, $v_a^{(2)}$ and $\dot{\mathbf{X}}^{(1)}$). However, in principle the derivation would proceed analogously to the case presented in section IV, with introduction of appropriate derivatives along stream lines defined by the second order continuity equation. 
Concretely, if $w_c^{(1)} \neq 0$ we can introduce the s-derivative along stream lines defined by the leading order axial mass flux, $\mathcal{M}^{(1)}_c$. This derivative is given  as
\begin{gather}
    D^{(1)}_s = \partial_s - \left(\mathcal{M}^{(1)}_c\right)_s\:/\:\left(\mathcal{M}^{(1)}_c\right)_{\bar{r}} 
    \partial_{\bar{r}} \qquad\qquad\qquad \\
    = \partial_s + \left(\frac{\sigma^{(0)}\bar{r}u_c^{(2)}}{w_c^{(1)}} +\frac{\bar{r}}{2}\frac{\left(\dot{\mathbf{X}}^{(0)}_s\cdot\hat{\tau}^{(0)}\right)}{w_c^{(1)}}-\frac{\sigma^{(0)}\kappa^{(0)}}{2w_c^{(1)}} \alpha z_b^{(0)} v^{(0)} I\right)\partial_{\bar{r}} \nonumber
\end{gather}
where we have used the superscript $(1)$ to distinguish this derivative from the $D_s$ derivative introduced in section IV. \\\\
This brief discussion highlights the importance of distinguishing between the three distinct cases of (1) non-vanishing leading order axial velocity ($w^{(0)} \neq 0$), (2) vanishing leading and first order axial velocity ($w^{(0)} = 0$ and $w^{(1)} = 0$) and (3) the intermediate case of vanishing leading order axial velocity ($w^{(0)} = 0$ and $w^{(1)} \neq 0$).  
\section{Impulse equation for a horizontal circular vortex ring}{Consider the case of a horizontally placed circular vortex ring of radius $R$ and core radius $\delta$ (as in Figure 2 B) with constant density and temperature within the core. We get
\begin{eqnarray}
C_T(s,t) 
&=& -8\pi^2 \int_0^{\infty}\xi\tilde{T}\; d\xi 
 = -8\pi^2 \int_0^{\delta}\xi\tilde{T}\; d\xi \nonumber\\
 &=& -8T\pi^2 \int_0^{\delta}\xi\; d\xi = -4\tilde{T}\pi^2\delta^2
\end{eqnarray}
So the part of the filament velocity resulting from the gravitational term, which we denote as $\dot{\mathbf{X}}_g^{(0)}$, is 
\begin{eqnarray}
\dot{\mathbf{X}}_g^{(0)}  
&=& \frac{1}{4\pi} \frac{g}{\Gamma}\beta  C_T(s,t) \hat{\mathbf{n}}^{(0)} \nonumber \\  &=& -\frac{1}{4\pi} \frac{g}{\Gamma}\beta 4\tilde{T}\pi^2\delta^2 \hat{\mathbf{n}}^{(0)}\nonumber \\
 &=& -\frac{g}{\Gamma}\beta \tilde{T}\pi \delta^2 \hat{\mathbf{n}}^{(0)}\nonumber \\  &=& -\frac{\pi \delta^2}{\Gamma} \beta \tilde{T} g \hat{\mathbf{n}}^{(0)}
\end{eqnarray}
So for a \textit{light ring} propagating upwards - that is $\Gamma \beta \tilde{T} > 0$ - the ring radius increases with time and the ring slows down. The absolute value of (D2) yields the rate of change of the vortex ring radius R. For such a vortex ring, \citet{Saffmann1993}(section 5.8) gives the hydrodynamic impulse $I=\rho_0 \Gamma R^2$ and  the buoyancy force $F_b = (\rho_0  - \rho_1 )2\pi g R \delta^2$ yielding the equilibrium equation 
$$\frac{dI}{dt} = 2 \rho_0 \Gamma R \frac{dR}{dt} = F_b$$ 
giving the rate of change of the vortex ring radius R, just as in (D2). Inserting $(1-\rho_1/\rho_0) = \beta \tilde{T}$ results in the same expression as that given in (D2).}

\bibliography{references}

\end{document}